\documentclass[11pt,letterpaper]{article}
\usepackage[dvips]{epsfig}
\usepackage{amsmath,amssymb,amsfonts,enumerate,bbm}
\usepackage{caption}
\usepackage{url}
\usepackage{fullpage}
\usepackage{ifthen}

\usepackage{graphicx}

\usepackage{subfigure}

\usepackage{color,soul}

\long\def\commabs #1\commabsend{}
\long\def\commful #1\commfulend{}
\long\def\comment #1\commentend{}

\newtheorem{theorem}{Theorem}[section]
\newtheorem{lemma}[theorem]{Lemma}
\newtheorem{observation}[theorem]{Observation}
\newtheorem{corollary}[theorem]{Corollary}

\newtheorem{fact}[theorem]{Fact}
\newtheorem{definition}[theorem]{Definition}

\newcommand{\REAL}{\mathbb R}
\def\deg{\mbox{\tt deg}}

\def\LCA{\mbox{\tt LCA}}
\def\DP{\mbox{\tt DP}}

\def\Cost{\mbox{\tt Cost}}

\newcommand{\dist}{\mbox{\rm dist}}

\def\inline#1:{\par\vskip 7pt\noindent{\bf #1:}\hskip 10pt}
\def\Proof{\par\noindent{\bf Proof:~}}
\def\blackslug{\hbox{\hskip 1pt \vrule width 4pt height 8pt
    depth 1.5pt \hskip 1pt}}
\def\QED{\quad\blackslug\lower 8.5pt\null\par}

\def\LastE{\mbox{\tt LastE}}

\def\source{s}
\def\sourceset{S}

\def\FTMBFS{\mbox{\tt FT-MBFS}}

\def\FTSMULTPANNERBFS{\mbox{\tt FT-AMBFS}}

\def\dnsparagraph#1{\par\vspace{2pt}\noindent{\bf #1}.}

\newcommand{\Conss}[0]{\mbox{\tt Cons4SWSpanner}}
\newcommand{\Consssix}[0]{\mbox{\tt Cons8SWSpanner}}

\def\Cost{\mbox{\tt Cost}}
\def\Value{\mbox{\tt Val}}
\def\ValSet{\mbox{\tt ValSet}}

\begin{document}



\title{Vertex Fault Tolerant Additive Spanners}

\author{
Merav Parter
\thanks{
Department of Computer Science and Applied Mathematics,
The Weizmann Institute of Science, Rehovot, Israel.
E-mail: {\tt \{merav.parter\}@ weizmann.ac.il}.
Recipient of the Google European Fellowship in distributed computing;
research is supported in part by this Fellowship.
Supported in part by the Israel Science Foundation (grant 894/09).}
}

\maketitle

\begin{abstract}
A {\em fault-tolerant} structure for a network is required to continue functioning following the failure of some of the network's edges or vertices.
In this paper, we address the problem of designing a {\em fault-tolerant} additive spanner, namely,
a subgraph $H$ of the network $G$
such that subsequent to the failure of a single vertex,
the surviving part of $H$ still contains
an \emph{additive} spanner for (the surviving part of) $G$, satisfying
$\dist(s,t,H\setminus \{v\}) \leq \dist(s,t,G\setminus \{v\})+\beta$ for every $s,t,v \in V$.
Recently, the problem of constructing fault-tolerant additive spanners resilient to the failure of up to $f$ \emph{edges} has been considered \cite{FTAdd12}. The problem of handling \emph{vertex} failures was left open therein.
In this paper we develop new techniques for constructing additive FT-spanners overcoming the failure of a single vertex in the graph. Our first result is an FT-spanner with additive stretch $2$ and $\widetilde{O}(n^{5/3})$ edges. Our second result is an FT-spanner with additive stretch $6$ and $\widetilde{O}(n^{3/2})$ edges. The construction algorithm consists of two main components: (a) constructing an FT-clustering graph and (b) applying a modified path-buying procedure suitably adopted to failure prone settings.
Finally, we also describe two constructions for {\em fault-tolerant multi-source additive spanners},
aiming to guarantee a bounded additive stretch following a vertex failure, for every pair of vertices in $S \times V$ for a given subset of sources $S\subseteq V$. The additive stretch bounds of our constructions are 4 and 8 (using a different number of edges).
\end{abstract}

\section{Introduction}
An $(\alpha,\beta)$-spanner $H$ of an unweighted undirected graph $G$ is a spanning subgraph satisfying for every pair of vertices $s,t \in V$ that $\dist(s,t,H) \leq \alpha \cdot \dist(s,t,G)+\beta$.
When $\beta=0$, the spanner is termed a \emph{multiplicative} spanner and when $\alpha=1$ the spanner is \emph{additive}. Clearly, additive spanners provide a much stronger guarantee than multiplicative ones, especially for long distances.
Constructions of additive spanners with \emph{small} number of edges are currently known for $\beta=2,4,6$ with $O(n^{3/2}), \widetilde{O}(n^{7/5})$ and $O(n^{4/3})$ edges respectively \cite{Aing99,New26,BSADD10,ChechikAdd13,DHZ00,PelegElkinMixedSpanner}.
This paper considers a network $G$ that may suffer single \emph{vertex} failure events, and looks for fault tolerant additive spanners that maintain their additive stretch guarantee under failures.
Formally, a subgraph $H \subseteq G$ is a $\beta$-additive FT-spanner iff for every $(s,t) \in V \times V$ and for every failing vertex $v \in V$,
$\dist(s,t, H \setminus \{v\}) \leq \dist(s,t, G \setminus \{v\})+\beta$.
As a motivation for such structures, consider a situation where it is required to lease a subnetwork of a given network, which will provide short routes from every source $s$ and every target $t$ with additive stretch $2$.
In a failure-free environment one can simply lease a 2-additive spanner $H_0$ of the graph with $\Theta(n^{3/2})$ edges.
However, if one of the vertices in the graph fails, some $s-t$ routes in $H_0 \setminus \{v\}$ might be significantly longer than the corresponding route in the  surviving graph $G \setminus \{v\}$. Moreover, $s$ and $t$ are not even guaranteed to be connected in $H_0 \setminus \{v\}$. One natural approach towards preparing for such eventuality is to lease a larger set of links, i.e., an additive FT-spanner.

The notion of fault-tolerant spanners for general graphs was initiated by Chechik at el. \cite{CLPR09-span} for the case of multiplicative stretch. Specifically, \cite{CLPR09-span} presented algorithms for constructing an $f$-vertex fault tolerant spanner with multiplicative stretch $(2k-1)$
and $O(f^2 k^{f+1} \cdot n^{1+1/k}\log^{1-1/k}n)$ edges.
Dinitz and Krauthgamer presented in \cite{DK11}, a randomized construction attaining an improved tradeoff for \emph{vertex} fault-tolerant spanners, namely, $f$-vertex fault tolerant $k$-spanner with $\widetilde{O}(f^2 \cdot n^{1+2/(k+1)})$ edges.
Constructions of fault-tolerant spanners with \emph{additive} stretch resilient to \emph{edge} failures were recently given by Braunschvig at el. \cite{FTAdd12}. They establish the following general result.
For a given $n$-vertex graph $G$, let $H_1$ be an ordinary additive $(1, \beta)$ spanner for $G$ and $H_2$ be a  $(\alpha,0)$  fault tolerant spanner for $G$ resilient against up to $f$ edge faults. Then $H=H_1 \cup H_2$ is a
$(1, \beta(f))$ additive fault tolerant spanner for $G$
(for up to $f$ edge faults) for $\beta(f)=O(f(\alpha+\beta))$.
In particular, fixing the number of $H$ edges to be $O(n^{3/2})$ and the number
of faults to $f=1$ yields an additive stretch of $14$. Hence, in particular, there is no construction for additive stretch $< 14$ and $o(n^2)$ edges.
In addition, note that these structures are resilient only to \emph{edge} failures as the techniques of \cite{FTAdd12} cannot be utilized to protect even against a single vertex failure event. Indeed, the problem of handling \emph{vertex} failures was left open therein.

In this paper, we make a first step in this direction and provide additive FT-structures resilient to the failure of a single \emph{vertex} (and hence also edge) event.
Our constructions provide additive stretch 2 and 6 and hence provide an improved alternative also
for the case of a single edge failure event, compared to the constructions of  \cite{FTAdd12}.

\par The presented algorithms are based upon two important notions, namely, \emph{replacement paths} and the \emph{path-buying procedure}, which have been studied extensively in the literature.
For a source $s$, a target vertex $t$ and a failing vertex $v\in V$, a \emph{replacement path} is the shortest $s-t$ path $P_{s,t,v}$ that does not go through $v$. The vast literature on \emph{replacement paths} (cf. \cite{BK09,GW12,RTREP05,WY10}) focuses on \emph{time-efficient} computation of the these paths as well as their efficient maintenance in data structures (a.k.a {\em distance oracles}).

Fault-resilient structures that preserve exact distances for a given subset of sources $S \subseteq V$ have been studied in \cite{PPFTBFS13}, which defines the notion of an \FTMBFS\ structure $H \subseteq G$ containing the collection of all replacement paths $P_{s,t,v}$ for every pair $(s,t) \in S \times V$ for a given subset of sources $S$ and a failing vertex $v \in V$.
Hence, \FTMBFS\ structures preserve the exact $s-t$ distances in $G \setminus \{v\}$ for every failing vertex $v$, for every source $s \in S$. 

It is shown in \cite{PPFTBFS13} that for every graph $G$ and a subset $S$ of sources, there exists a (poly-time constructible) 1-edge (or vertex) \FTMBFS\ structure $H$ with
$O(\sqrt{|S|} \cdot n^{3/2})$ edges. This result is complemented by a matching lower bound
showing that for
sufficiently large
$n$, there exist an $n$-vertex graph $G$ and a source-set $S \subseteq V$, for which every \FTMBFS\
structure is of size $\Omega(\sqrt{|S|} \cdot n^{3/2})$.
Hence {\em exact} \FTMBFS\ structures may be rather expensive. This last observation motivates the approach of resorting to \emph{approximate}
distances, in order to allow the design of a sparse subgraph with properties resembling those of an \FTMBFS\ structure.

The problem of constructing
\emph{multiplicative approximation replacement paths} $\widetilde{P}_{s,t,v}$
(i.e., such that $|\widetilde{P}_{s,t,v}| \leq \alpha \cdot |P_{s,t,v}|$) has been studied in \cite{BS06,CLPR09-do,Bern10}.
In particular its {\em single source} variant has been studied in \cite{BS10,PPFTBFS14}.
In this paper, we further explore this approach. For a given subset of sources $S$, we focus on constructions of subgraphs that contain an {\em approximate} BFS
structure with additive stretch $\beta$ for every source $s \in S$ that are resistant to a single vertex failure.

Indeed, the construction of additive sourcewise FT-spanners provides a key building block of additive FT-spanner constructions (in which bounded stretch is guaranteed for all pairs).  
%
We present two constructions of sourcewise spanners with different stretch-size tradeoffs. The first construction ensures an additive stretch $4$ with $\widetilde{O}(\max\{ n \cdot |S|, (n/|S|)^3\})$ edges and the second construction guarantees additive stretch $8$ with $\widetilde{O}(\max\{ n \cdot |S|, (n/|S|)^2\})$. As a direct consequence of these constructions, we get an additive FT-spanner with stretch $6$ and $\widetilde{O}(n^{3/2})$ edges and an additive sourcewise FT-spanner with additive stretch $8$ and $\widetilde{O}(n^{4/3})$ for at most $\widetilde{O}(n^{1/3})$ sources.

Our constructions employ a modification of the \emph{path-buying} strategy, which was originally devised in \cite{BSADD10} to provide $6$-additive spanners with $O(n^{4/3})$ edges. Recently, the path-buying strategy was employed in the context of pairwise spanners, where the objective is to construct a subgraph $H \subseteq G$ that satisfies the bounded additive stretch requirement only for a \emph{subset} of pairs \cite{CGK13}.
The high-level idea of this procedure as follows. In an initial clustering phase, a suitable clustering of the vertices is computed, and an associated subset of edges is added to the spanner. Then comes a path-buying phase, where they consider an appropriate sequence of paths, and decide whether or not to add each path into the spanner.
Each path $P$ has a {\em cost},
given by the number of edges of $p$ not already contained in the spanner, and a {\em value},
measuring $P$'s help in satisfying the considered set of constraints on pairwise distances.
The considered path $P$ is added to the spanner iff its value is sufficiently larger than its cost.
In our adaptation to the FT-setting, an FT-clustering graph is computed first, providing every vertex with a sufficiently \emph{high} degree (termed hereafter a heavy vertex) \emph{two} clusters to which it belongs. Every cluster consists of a center vertex $v$ connected via a star to a subset of its heavy neighbors. In our design not all replacement paths are candidates to be bought in the path-buying procedure. Let $\pi(s,t)$ be an $s-t$ shortest-path between a source $s$ and a heavy vertex $t$ (in our constructions, all heavy vertices are clustered). We divide the failing events on $\pi(s,t)$ into two classes depending on the position of the failing vertex on $\pi(s,t)$ with respect to the least common ancestor (LCA) $\ell(s,t)$ of $t$'s cluster members in the BFS tree rooted at $s$. Specifically, a vertex fault $\pi(s,t)$ that occurs on $\ell(s,t)$ is handled \emph{directly} by adding the last edge of the corresponding replacement path to the spanner.
Vertex failures that occur strictly below the LCA, use the shortest-path $\pi(s,x)$ between $s$ and some member $x$ in the cluster of $t$ whose failing vertex $v$ does not appear on its $\pi(s,x)$ path. The approximate replacement path will follow $\pi(s,x)$ and then use the intercluster path between $x$ and $t$. The main technicality is when concerning the complementary case when that failing events occur strictly above $\ell(s,t)$. These events are further divided into two classes depending on the structure of their replacement path.
Some of these replacement paths would again be handled directly by collecting their last edges into the structure and only the second type paths would be candidate to be bought by the path-buying procedure. Essentially, the structure of these paths and the cost and value functions assigned to them would guarantee that the resulting structure is sparse, and in addition, that paths that were not bought have an alternative safe path in the surviving part of the structure.

\vspace{-3pt}
\paragraph{Contributions.}
This paper provides the first constructions for additive spanners resilient upon single vertex failure. In addition, it provides the first additive FT-structures with stretch guarantee as low as 2 or 6 and with $o(n^2)$ edges.

The main technical contribution of our algorithms is in adapting the path-buying
strategy to the vertex failure setting.
Such an adaptation has been initiated in \cite{PPFTBFS14}
for the case of a \emph{single-source} $s$ and a single \emph{edge} failure event. In this paper, we extend this technique in two senses: (1) dealing with many sources and (2) dealing with \emph{vertex} failures.
In particular, \cite{PPFTBFS14} achieves a construction of single source additive spanner with $O(n^{4/3})$ edges resilient to a single \emph{edge} failure. In this paper, we extend this construction to provide a multiple source additive spanners resilient to a single vertex failure, for $O(n^{1/3})$ sources, additive stretch 8 and $\widetilde{O}(n^{4/3})$ edges.
In summary, we show the following.
\begin{theorem}[2-additive FT-spanner]
\label{thm:2stretch}
For every
$n$-vertex graph $G=(V,E)$, there exists a (polynomially constructible) subgraph $H \subseteq G$ of size $\widetilde{O}(n^{5/3})$ such that $\dist(s,t,H\setminus\{v\})\leq \dist(s,t,G\setminus\{v\})+2$ for every
$s,t,v \in V$.
\end{theorem}

\begin{theorem}[6-additive FT-spanner]
\label{thm:6stretch}
For every
$n$-vertex graph $G=(V,E)$, there exists a (polynomially constructible) subgraph $H \subseteq G$ of size $\widetilde{O}(n^{3/2})$ such that $\dist(s,t,H\setminus\{v\})\leq \dist(s,t,G\setminus\{v\})+6$ for every
$s,t,v \in V$.
\end{theorem}

\begin{theorem}[8-additive sourcewise FT-spanner]
\label{thm:8sw}
For every
$n$-vertex graph $G=(V,E)$ and a subset of sources $S \subset V$ where $|S|=\widetilde{O}(n^{1/3})$, there exists a (polynomially constructible) subgraph $H \subseteq G$ of size $\widetilde{O}(n^{4/3})$ such that $\dist(s,t,H\setminus\{v\})\leq \dist(s,t,G\setminus\{v\})+8$ for every
$s\in S$ and $t,v \in V$.
\end{theorem}
\section{Preliminaries}
\paragraph{Notation.}
Given a graph $G=(V,E)$, a vertex pair $s,t$ and
an edge weight function $W: E(G)\to \REAL^{+}$, let $SP(s, t, G, W)$ be the set of $s-t$ shortest-paths in $G$ according to the edge weights of $W$.
Throughout, we make use of (an arbitrarily specified) weight assignment $W$ that guarantees the uniqueness of the shortest paths\footnote{The role of the weights $W$ is to perturb the edge weights by letting $W(e)=1+\epsilon$ for a random infinitesimal $\epsilon>0$.}. Hence, $SP(s, t, G', W)$ contains a single path for every $s,t \in V$ and for every subgraph $G' \subseteq G$, we override notation and let $SP(s, t, G, W)$ be the unique $s-t$ path in $G$ according to $W$.
When the shortest-path are computed in $G$, let $\pi(s,t)=SP(s, t, G, W)$.
To avoid cumbersome notation, we may omit $W$ and simply refer to $\pi(s,t)=SP(s, t, G, W)$.
For a subgraph $G' \subseteq G$, let $V(G')$ (resp., $E(G')$) denote the vertex set (resp. edge set) in $G'$.
\par For a given source node $s$, let $T_0(s)=\bigcup_{t \in V} \pi(s,t)$ be a shortest paths (or BFS) tree rooted at $s$. For a set $S \subseteq V$ of source nodes,
let $T_0(S)=\bigcup_{s \in S} T_0(s)$ be a union of the single source BFS trees. For a vertex $t \in V$ and a subset of vertices $V' \in V$, let $T(t,V')=\bigcup_{u \in V'} \pi(u,t)$ be the union of all $\{t\} \times V'$ shortest-paths (by the uniqueness of $W$, $T(t,V')$ is a subtree of $T_0(t)$).
Let $\Gamma(v, G)$ be the set of $v$'s neighbors in $G$. Let $E(v,G)=\{(u,v) \in E(G)\}$ be the set of edges incident to $v$
in the graph $G$ and let $\deg(v,G)=|E(v,G)|$ denote the degree of node $v$ in $G$. 
%
For a given graph $G=(V,E)$ and an integer $\Delta \leq n$, a vertex $v$ is $\Delta$-\emph{heavy} if $\deg(v,G) \geq \Delta$, otherwise it is $\Delta$-\emph{light}. When $\Delta$ is clear from the context, we may omit it and simply refer to $v$ as \emph{heavy} or \emph{light}.
For a graph $G=(V,E)$ and a positive integer $\Delta \leq n$, let $V_{\Delta}=\{v ~\mid~ \deg(v,G)\geq \Delta\}$ be the set of $\Delta$-heavy vertices in $G$.
(Throughout, we sometimes simplify notation by omitting parameters which are clear from the context.)
For a subgraph $G'=(V', E') \subseteq G$
(where $V' \subseteq V$ and $E' \subseteq E$)
and a pair of vertices $u,v \in V$, let $\dist(u,v, G')$ denote the
shortest-path distance in edges between $u$ and $v$ in $G'$.
For a path $P=[v_1, \ldots, v_k]$, let $\LastE(P)$ be the last edge of $P$, let $|P|$ denote its length and let $P[v_i, v_j]$ be the subpath of $P$ from $v_i$ to $v_j$. For paths $P_1$ and $P_2$,  denote by $P_1 \circ P_2$ the path obtained by concatenating $P_2$ to $P_1$.
For ``visual" clarity, the edges of these paths are considered throughout, to be directed away from the source node $s$.
Given an $s-t$ path $P$ and an edge $e=(x,y) \in P$, let $\dist(s, e, P)$ be the distance (in edges) between $s$ and $y$ on $P$. In addition, for an edge $e=(x,y)\in T_0(s)$, define $\dist(s,e)=i$ if $\dist(s,x,G)=i-1$ and $\dist(s,y,G)=i$.
A vertex $w$ is a \emph{divergence point} of the $s-v$ paths  $P_1$ and $P_2$ if $w \in P_1 \cap P_2$ but the next vertex $u$ after $w$ (i.e., such that $u$ is closer to $v$) in the path $P_1$ is not in $P_2$.
\paragraph{Basic Tools.}
We consider the following graph structures.
\begin{definition}[$(\alpha,\beta,S)$-AMBFS FT-spanners]
A subgraph $H \subseteq G$ is
an $(\alpha, \beta,S)$-\FTSMULTPANNERBFS\ (approximate multi-BFS) structure with respect to $S$ if for every $(s,t) \in S \times V$ and every $v \in V$,
$\dist(s, t, H \setminus \{v\}) \leq
\alpha \cdot \dist(s, t, G \setminus \{v\})+\beta~.$
\end{definition}

\begin{definition}[$(\alpha, \beta)$ FT-spanners]
A subgraph $H \subseteq G$ is an $(\alpha, \beta)$ FT-spanner if it is an $(\alpha, \beta,V)$-\FTSMULTPANNERBFS\ structure for $G$ with respect to $V$.
\end{definition}
Throughout, we restrict attention to the case of a single vertex fault. When $\alpha=1$, $H$ is termed  $(\beta, S)-$ additive FT-spanner. In addition, in case where $S=V$, $H$ is an $\beta$-additive FT-spanner.

\paragraph{FT-Clustering Graph.}
A subset $Z \subseteq V$ is an \emph{FT-center} set for $V$ if every $\Delta$-\emph{heavy} vertex $v$ has at least two neighbors in $Z$, i.e., $|\Gamma(v,G) \cap Z|\geq 2$.
For every heavy vertex $v \in V_{\Delta}$, let $Z(v)=\{z_1(v), z_2(v)\}$ be two arbitrary neighbors of $v$ in $Z$. The clustering graph $G_{\Delta} \subseteq G$ consists of the edges connecting the $\Delta$-heavy vertices $v$ to their two representatives in $Z$ as well as all edges incident to the $\Delta$-light vertices. Formally, $$G_\Delta=\bigcup_{v \in V_{\Delta}}\{(v,z_1(v)), (v,z_2(v))\} \cup \bigcup_{v \notin  V_{\Delta}} E(v,G).$$

The $\Delta$-\emph{heavy} vertices are referred hereafter as \emph{clustered}, hence every missing edge in $G \setminus G_{\Delta}$ is incident to a clustered vertex.

For every center vertex $z \in Z$, let $C_z$ be the cluster consisting of $z$ and all the $\Delta$-heavy vertices it represents, i.e., $C_z=\{z\} \cup \{v \in V_{\Delta}~\mid~ z \in Z(v)\}.$
Note that every center $z$ is connected via a star to each of the vertices in its cluster $C_z$, hence the diameter of each cluster $C_z$ in $G_{\Delta}$ is $2$.

For a failing vertex $v$ and a heavy vertex $t$, let $z_v(t) \in Z(t) \setminus \{v\}$ be a cluster center of $t$ in $G\setminus \{v\}$. In particular, if $z_1(t)\neq v$, then
$z_v(t)=z_1(t)$, else $z_v(t)=z_2(t)$. Let $C_v(t)$ be the cluster centered at $z_v(t)$.
Note that since every heavy vertex has two cluster centers $z_1(t)$ and $z_2(t)$, we have the guarantee that at least one of them survives the single vertex fault event.
The next observation summarizes some important properties of the clustering graph.
\begin{observation}
\label{obs:clust_prop}
(1) $|E(G_{\Delta})|=O(\Delta \cdot n)$.\\
(2) Every missing edge is incident to a clustered vertex in $V_{\Delta}$.\\
(3) The diameter of every cluster $C_z$ is $2$.\\
(4) There exists an FT-center set $Z \subseteq V$ of size $|Z|=\widetilde{O}(n/\Delta)$.
\end{observation}
Obs. \ref{obs:clust_prop}(4) follows by a standard hitting set argument.
%
%
%
\paragraph{Replacement Paths.}
For a source $s$, a target vertex $t$ and a vertex $v\in G$,
a \emph{replacement path} is the shortest $s-t$ path $P_{s,t,v} \in SP(s, t, G \setminus \{v\})$ that does not
go through $v$.

\begin{observation}
\label{obs:nmissing}
Every path $P_{s,t,v}$ contains at most $3n/\Delta$ $\Delta$-heavy vertices.
\end{observation}
\Proof
Note that
\begin{eqnarray*}
3n ~\geq~ 3 \cdot \left|\bigcup_{x \in P_{s,t,v} \cap V_{\Delta}}\Gamma(x,G\setminus\{v\})\right|\geq
\sum_{x \in P_{s,t,v} \cap V_{\Delta}}\deg(x,G\setminus\{v\})
\geq
|P_{s,t,v} \cap V_\Delta| \cdot \Delta~,
\end{eqnarray*}
where the second inequality follows by the fact the every vertex $u \in V \setminus \{v\}$ has at most $3$ neighbors on $P_{s,t,v}$. The observation follows.
\QED

\dnsparagraph{New-ending replacement paths}
A replacement path $P_{s,t,v}$ is called \emph{new-ending} if its last edge is different from the last edge of the shortest path $\pi(s,t)$. Put another way, a new-ending replacement path $P_{s,t,v}$ has the property that once it diverges from the shortest-path $\pi(s,t)$ at the vertex $b$, it joins $\pi(s,t)$ again only at the final vertex $t$.
It is shown in \cite{PPFTBFS13} that for a given graph $G$ and a set $S$ of source vertices, a structure $H \subseteq G$ containing a BFS tree rooted at each $s \in S$ plus the last edge of each new-ending replacement path $P_{s,t,v}$ for every $(s,t) \in S \times V$ and every $v \in V$, is an \FTMBFS\ structure with respect to $S$.
Our algorithms exploit the structure of new-ending replacement paths to construct $(\beta,S)$-additive FT-spanners. Essentially, a key section in our analysis concerns with collecting the last edges from a subset of new-ending replacement paths as well as bounding the number of new-ending paths $P_{s,t,v}$ whose detour segments intersect with $\pi(s',t)\setminus \{t\}$ for some other source $s' \in S$.
\paragraph{The basic building block.}
Our constructions of $\beta$-additive FT-spanners, for $\beta\geq 2$, consist of the following two building blocks: (1) an FT-clustering graph $G_{\Delta}$ for some parameter $\Delta$, and (2) an  $(\beta-2,Z)$-additive FT-spanner where $Z$ is an FT-center set (i.e., cluster centers) for the vertices.
\begin{lemma}
\label{obs:ftsources}
Let $\beta\geq 2$ and $H=G_{\Delta} \cup H_{\beta-2}(Z)$ where $Z$ is an FT-center set for $V_{\Delta}$. Then $H$ is an $\beta$ additive FT-spanner.
\end{lemma}
\Proof
Consider vertices $u_1,u_2,u_3 \in V$.
Let $P \in SP(u_1,u_2, G \setminus \{u_3\})$ be the $u_1-u_2$ replacement path in $G \setminus \{u_3\}$ and let $(x,y)$ be the last missing edge on $P \setminus H$ (i.e., closest to $u_2$). Since $G_{\Delta} \subseteq H$, by Obs. \ref{obs:clust_prop}(2), $y$ is a clustered vertex.
Let $z =z_{u_3}(y)$ be the cluster center of $y$ in $G \setminus \{u_3\}$, and consider the following $u_1-u_2$ path $P_3=P_1 \circ P_2$ where $P_1 \in SP(u_1, z, H \setminus \{u_3\})$ and $P_2 =(z,y) \circ P[y, u_2]$.
Clearly, $P_3 \subseteq H \setminus \{u_3\}$, so it remains to bound its length. Since $H_{\beta-2}(Z) \subseteq H$, it holds that $|P_1|\leq \dist(u_1, z, G \setminus \{u_3\})+\beta-2$. Hence,
\begin{eqnarray*}
\dist(u_1, u_2, H \setminus \{u_3\}) &\leq& |P_3|~=~|P_1|+|P_2|
\\&\leq&
\dist(u_1,z, G \setminus \{u_3\})+\beta-2+\dist(y,u_2, G \setminus \{u_3\})
\\&\leq&
\dist(u_1,y, G \setminus \{u_3\})+\dist(y, u_2, G \setminus \{u_3\})+\beta
\\&\leq&
|P|+\beta~=~\dist(u_1, u_2, G \setminus \{u_3\})+\beta~,
\end{eqnarray*}
where the second inequality follows by the triangle inequality using the fact that the edge $(z,y)$ exists in $H \setminus \{u_3\}$. The lemma follows.
\QED
\section{Additive Stretch 2}
\label{sec:2add}

We begin by considering the case of additive stretch $2$. We make use of the construction of \FTMBFS\ structures presented in \cite{PPFTBFS13}.

\begin{fact}[\cite{PPFTBFS13}]
\label{thm:multi_source_edgeonef}
There exists a polynomial time algorithm that for every $n$-vertex graph $G=(V,E)$ and source set $S \subseteq V$ constructs an \FTMBFS\ structure $H_{0}(S)$ from each source $s_i \in S$, tolerant to one edge or vertex failure, with a total number of $O(\sqrt{|S|} \cdot n^{3/2})$ edges.
\end{fact}
Set $\Delta=\lceil n^{2/3} \rceil $ and let $Z$ be an FT-center set for $V_{\Delta}$ as given by Obs. \ref{obs:clust_prop}(4). Let $H_{0}(Z)$ be an \FTMBFS\ structure with respect to the source set $Z$ as given by Fact \ref{thm:multi_source_edgeonef}.
Then, let
$H=G_\Delta \cup H_{0}(Z).$
Thm. \ref{thm:2stretch} follows by Lemma \ref{obs:ftsources}, Obs. \ref{obs:clust_prop} and Fact \ref{thm:multi_source_edgeonef}.
\section{Sourcewise additive FT-spanners}
\label{sec:sourcewise}
In this section, we present two constructions of
$(4,S)$ and $(8,S)$ additive FT-spanners with respect to a given source set $S \subseteq V$.
The single source case (where $|S|=1$) is considered in \cite{PPFTBFS14}, which provides a construction of a single source FT-spanner\footnote{The construction of \cite{PPFTBFS14} supports a single edge failure, yet, it can be modified to overcome a single vertex failure as well.} with $O(n^{4/3})$ edges and additive stretch $4$. The current construction increases the stretch to 8 to provide a bounded stretch for $\widetilde{O}(n^{1/3})$ sources with the same order of edges, $\widetilde{O}(n^{4/3})$.
\subsection{Sourcewise spanner with additive stretch 4}
\begin{lemma}
\label{lem:h4}
There exists a subgraph $H_{4}(\sourceset) \subseteq G$ with $\widetilde{O}(\max\{|\sourceset| \cdot n, (n/|\sourceset|)^{3}\})$ edges satisfying
$\dist(\source,t, H_{4}(\sourceset) \setminus \{v\})\leq \dist(\source,t,G \setminus \{v\})+4$ for every $(\source,t) \in \sourceset \times V$ and $v \in V$.
\end{lemma}

The following notation is useful in our context.
Let $\mathcal{C}=\{C_z ~\mid~ z \in Z\}$ be the collection of clusters corresponding to the FT-centers $Z$.
For a source $\source \in \sourceset$ and a cluster $C_z \in \mathcal{C}$ rooted at FT-center $z \in Z$, let $\LCA(\source,C_z)$ be the least common ancestor (LCA) of the cluster vertices of $C_z$ in the BFS tree $T_0(\source)$ rooted at $\source$.
Let $\pi(\source,C_z)$ be the path connecting $\source$ and $\LCA(\source,C_z)$ in $T_0(\source)$.

\subsubsection{Algorithm \Conss\ for constructing $H_4(\sourceset)$ spanner}
\paragraph{Step (0): Replacement-path definition.}
For every $(s,t) \in S \times V$ and every $v \in V$, let $P_{s,t,v}=SP(s,t, G \setminus \{v\},W)$.
\paragraph{Step (1): Clustering.}
Set $\Delta=|S|$ and let $Z \subseteq V$ be an
FT-center set of size $\widetilde{O}(n/\Delta)$ (by Obs. \ref{obs:clust_prop}(4) such set exists). Let $\mathcal{C}=\{C_z ~\mid~ z \in Z\}$ be the collection of $|Z|$ clusters.
For a heavy vertex $t$, let $C_1(t), C_2(t)$ be its two clusters in $\mathcal{C}$ corresponding to the centers $z_1(t)$ and $z_2(t)$ respectively.

\paragraph{Step (2): Shortest-path segmentation.}
For every $(s,t) \in S \times V_{\Delta}$, the algorithm uses the first cluster of $t$, $C_1(t)$, to
segment
the path $\pi(\source,t)$.
Define
$$\pi^{far}(\source,t)=\pi(\source,\ell(s,t)) \setminus \{\ell(s,t)\} \mbox{~~and~~} \pi^{near}(\source,t)=\pi(\ell(s,t),t)\setminus \{\ell(s,t)\},$$
where $\ell(s,t)=\LCA(\source,C_1(t))$ is the LCA of the cluster $C_1(t)$ in the tree $T_0(s)$.
Hence, $\pi(s,t)=\pi^{far}(\source,t) \circ \ell(s,t) \circ \pi^{near}(\source,t)$.
The algorithm handles separately vertex faults in the near and far segments. Let
$V^{near}(s,t)=V(\pi^{near}(\source,t))$ and $V^{far}(s,t)=V(\pi^{far}(\source,t))$. 
\paragraph{Step (3): Handling faults in the cluster center and the LCA.}
Let
$$E^{local}(t)=\{ \LastE(P_{s,t,v}) ~\mid~ s \in S, v  \in \{z_1(t),\LCA(s,C_1(t))\}\} \mbox{~and~}
E^{local}=\bigcup_{t \in V_{\Delta}} E^{local}(t),$$
be the last edges of replacement-paths protecting against the failure of the primary cluster center $z_1(t)$ and the least common ancestor $\LCA(s,C_1(t))$.
\paragraph{Step (4): Handling far vertex faults $V^{far}(s,t)$.}
A replacement path $P_{\source,t,v}$ is \emph{new-ending} if its last edge is not in $(T_0(S) \cup G_{\Delta})$. For a new-ending path $P_{\source,t,v}$, let $b_{s,t,v}$ be the unique divergence point of $P_{s,t,v}$ from $\pi(s,t)$ (in the analysis we show that such point exists).
Let $D_{\source,t,v}=P_{\source,t,v}[b_{s,t,v},t]$
denote the detour segment and let $D^+_{\source,t,v}=D_{\source,t,v}\setminus \{b_{s,t,v}\}$ denote the detour segment excluding the divergence point.
For every clustered vertex $t$, let $\mathcal{P}^{far}(t)$ be the collection of new-ending $\source-t$ paths protecting against vertex faults in the far segments, i.e.,
$\mathcal{P}^{far}(t)=\{P_{\source,t,v} ~\mid~ \source \in \sourceset, ~ \LastE(P_{\source,t,v}) \notin T_0(\sourceset) \mbox{~and~} v \in V^{far}(s,t)\}$.

\begin{figure}[h!]
\begin{center}
\includegraphics[scale=0.4]{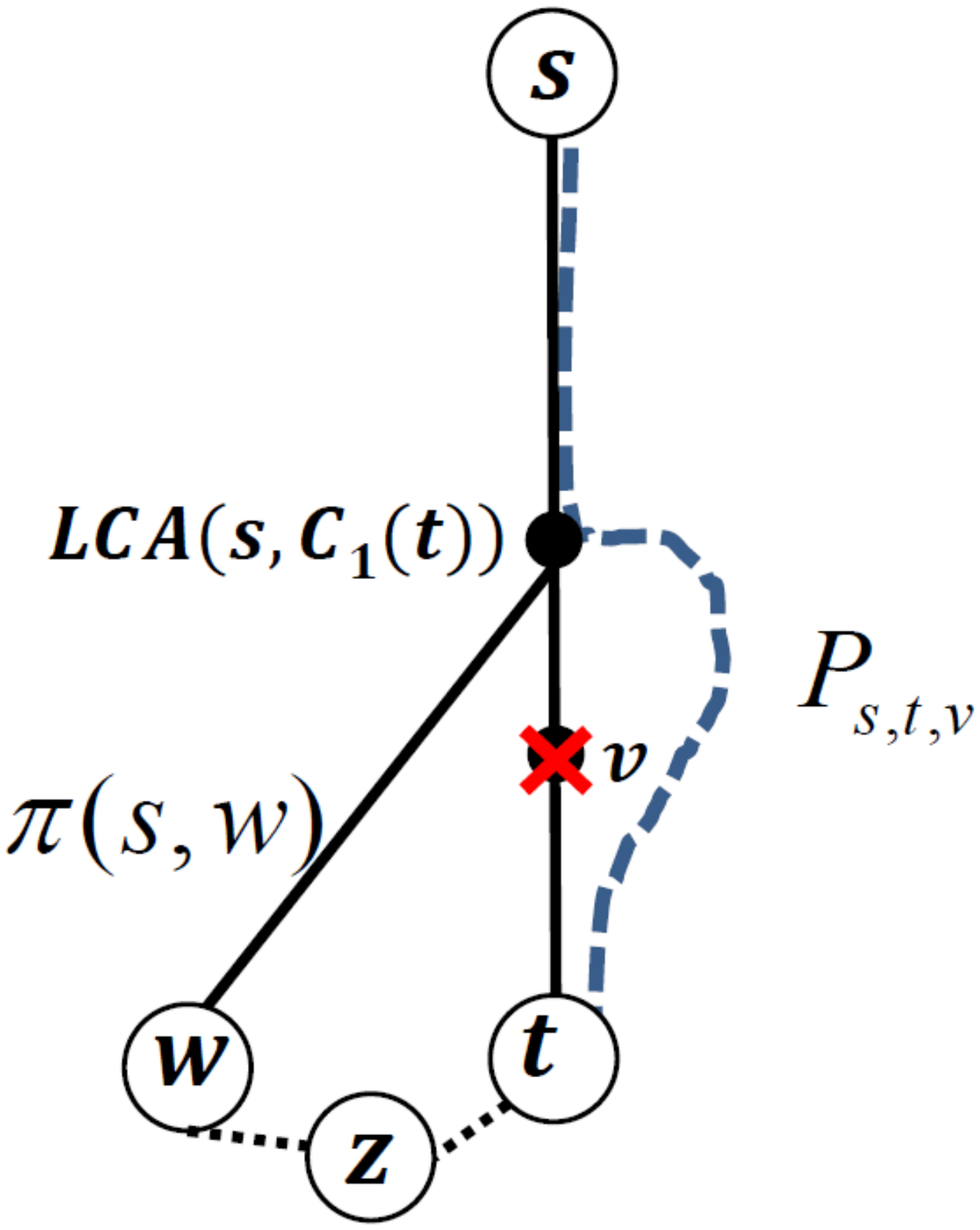}
\caption{Handling near vertex faults.
Schematic illustration of an approximate replacement path in $(T_0 \cup G_\Delta)\setminus \{v\}$. Shown is an $\pi(s,t)$ whose failing vertex $v$ occurs strictly below the least common ancestor $\LCA(s, C_1(t))$. The alternative replacement path exploits the surviving $\pi(s,w) \subseteq T_0(S)$ path for $w \in C_1(t)$ and the intracluster path connecting $w$ and $v$ through $z=z_v(t)$.
\label{fig:near}}
\end{center}
\end{figure}

The algorithm divides this set into two subsets $\mathcal{P}^{far}_{dep}(t)$ and $\mathcal{P}^{far}_{indep}(t)$ depending on the structure of the partial detour segment $D^+_{\source,t,v}$.
A new-ending path $P_{\source,t,v}$ is \emph{dependent} if $D^+_{\source,t,v}$ intersects $\pi(\source',t)\setminus\{t\}$ for some $\source' \in \sourceset$, i.e., for a dependent path $P_{s,t,v}$, it holds that
\begin{equation}
\label{eq:depend}
V(D^+_{\source,t,v}) \cap V(T(t,\sourceset)) \neq \{t\}~.
\end{equation}
Otherwise, it is \emph{independent}.
Let $$\mathcal{P}^{far}_{dep}(t)=\{P_{\source,t,v} \in \mathcal{P}^{far}(t) ~\mid~ s \in S, v \in V^{far}(s,t) \mbox{~and~} V(D^+_{\source,t,v}) \cap V(T(t,\sourceset))\neq  \{t\}\}$$ be the set of all $S \times \{t\}$ dependent paths and let $\mathcal{P}^{far}_{indep}(t)=\mathcal{P}^{far} \setminus \mathcal{P}^{far}_{dep}(t)$ be the set of independent paths.
\paragraph{Step (4.1): Handling \emph{dependent} new-ending paths.}
The algorithm simply takes the last edges $E^{far}_{dep}(t)$ of all dependent replacement paths where $E^{far}_{dep}(t)=\{\LastE(P) ~\mid~ P \in \mathcal{P}^{far}_{dep}(t)\}$.
(In the analysis section, we show that the $E^{far}_{dep}(t)$ sets are sparse.)
Let $E^{far}_{dep}=\bigcup_{t \in  V_{\Delta}} E^{far}_{dep}(t)$.
\paragraph{Step (4.2): Handling \emph{independent} new-ending paths.}
The algorithm employs a modified \emph{path-buying} procedure on the collection
$\mathcal{P}^{far}_{indep}=\bigcup_{t \in V_{\Delta}}\mathcal{P}^{far}_{indep}(t)$ of new-ending independent paths.
The paths of $\mathcal{P}^{far}_{indep}$ are considered in some arbitrary order.
A path $P \in \mathcal{P}^{far}_{indep}$ is bought, if it improves the pairwise cluster distances in some sense.
Starting with
\begin{equation}
\label{eq:g0}
G_0=T_0(\sourceset) \cup G_{\Delta} \cup E^{local} \cup E^{far}_{dep}~,
\end{equation}
at step $\tau \geq 0$, the algorithm is given $G_{\tau} \subseteq G$ and considers the path $P_{\tau}=P_{\source,t, v}$.
Let $e=(x,y)$ be the first missing edge on $P_{\tau} \setminus E(G_\tau)$ (where $x$ is closer to $\source$).
Note that since $G_{\Delta} \subseteq G_0$, both $x$ and $t$ are clustered. Recall that for a clustered vertex $u$ and a failing vertex $v$, $C_v(u)$ is the cluster of $u$ centered at $z_v(u) \in Z(u)\setminus \{v\}$.
For every cluster $C$, let $V_{f}(C)$ be the collection of vertices appearing on the paths $\pi(\source,C)=\pi(s, \LCA(s,C))$
for every $\source\in \sourceset$ excluding the vertices of the clusters. That is,
\begin{equation}
\label{eq:hctau}
V_{f}(C)=\bigcup_{\source \in \sourceset} V(\pi(\source, C))\setminus C.
\end{equation}
%
%
%
%
%
The path $P_\tau$ is added to $G_\tau$ resulting in $G_{\tau+1}=G_{\tau} \cup P_{\tau}$,
only if
\begin{equation}
\label{eq:pathbuying6add}
\dist(x, t, P_\tau)< \dist(C_v(x), C_v(t), G_\tau \setminus V_{f}(C_v(t))).
\end{equation}
Let $\tau'=|\mathcal{P}^{far}_{indep}|$ be the total number of independent paths considered to be bought by the algorithm.
Then, the algorithm outputs
$H_4(\sourceset)=G_{\tau'}.$
This completes the description of the algorithm.
\paragraph{Analysis.}
Throughout the discussion, we consider a $P_{s,t,v}$ paths of clustered vertices $t \in V_{\Delta}$. A path $P_{s,t,v}$ is a new-ending path, if $\LastE(P_{s,t,v})\notin G_0$ (see Eq. (\ref{eq:g0})).
Let $b_{s,t,v}$ be the first divergence point of $P_{s,t,v}$ and $\pi(s,t)$.
\begin{lemma}
\label{cl:new}
For every vertex $u \in P_{s,t,v}$ such that $\LastE(P_{s,t,v}[s,u]) \notin T_0(S)$, it holds that:
(a) $v \in \pi(s,u)$.~~
(b) $V(P_{s,t,v}[b_{s,t,v},u]) \cap V(\pi(s,u))=\{b_{s,t,v},u\}$.
\end{lemma}
\Proof
Begin with (a).
Assume towards contradiction otherwise.
By the uniqueness of the weight assignment $W$, we get that
$P_{s,t,v}[s,u]=SP(s,u, G \setminus \{v\},W)=\pi(s,u)$.
Leading to contradiction to the fact that $\LastE(P_{s,t,v})$ not in $T_0(S)$.
We next prove (b) and show that the divergence point $b_{s,t,v}$ is unique.
By the definition of $b_{s,t,v}$, it occurs on $\pi(s,t)$ above the failing vertex $v$.
Since by Lemma \ref{cl:new}, $v \in \pi(s,u)$, it also holds that $b_{s,t,v} \in \pi(s,u)$. Assume towards contradiction that there exists an additional point
$$w \in  \left(P_{s,t,v}[b_{s,t,v},u] \cap \pi(s,u)\right) \setminus \{b_{s,t,v},u\}.$$
There are two cases to consider (b1) $v \in \pi(b_{s,t,v},w)$, in such a case, $v \notin \pi(w,u)$ and hence $\pi(w,u)=SP(w,u, G \setminus \{v\})=P_{s,t,v}[w,u]$, contradiction that $\LastE(P_{s,t,v}[s,u]) \notin T_0(S)$.
(b2) $v \in \pi(w,u)$. In such a case, $v \notin \pi(b_{s,t,v},w)$ and hence $\pi(b_{s,t,v},w)=SP(b_{s,t,v},w, G \setminus \{v\})= P_{s,t,v}[b_{s,t,v},w]$, contradiction to the fact the $b_{s,t,v}$ is a divergence point from $\pi(s,t)$. The claim holds.
\QED
%
The next claim shows that a new-ending $P_{s,t,v}$ path whose last edge is not in $G_0$ (see Eq. (\ref{eq:g0})), protecting against faults in the near segment, has a good approximate replacement $\widetilde{P}_{s,t,v}$ in $T_0 \cup G_\Delta$.
\begin{lemma}
\label{lem:near}
If $\LastE(P_{s,t,v}) \notin G_0$ and $v \in \pi^{near}(s,t)$,  then
$\dist(s,t, (G_0 \cup G_{\Delta}) \setminus \{v\})\leq \dist(s,t, G\setminus \{v\})+4$.
\end{lemma}
\Proof
Since $v \in \pi^{near}(s,t)$, i.e., the failing vertex occurs strictly below $\LCA(s, C_1(t))$ on $\pi(s,t)$, there exists a vertex $w \in C_1(t)$ such that $v \notin \pi(s,w)$ (hence in particular $w \neq v$).
See Fig. \ref{fig:near}.
Since $\LastE(P_{s,t,v}) \notin E^{local}$, it holds that $v \neq z_1(t)$.
Consider the following $s-w$ path $P=\pi(s,w) \circ [w, z_1(t),t]$. Clearly, $P \subseteq (T_0(S) \cup G_{\Delta}) \setminus \{v\}$. By the triangle inequality, as the diameter of the cluster $C_1(t)$ is $2$, it holds that
\begin{eqnarray*}
\dist(s,t, (T_0(S) \cup G_{\Delta}) \setminus \{v\})~\leq~ \dist(s,w,G)+2 \leq
\dist(s,t,G)+4 \leq \dist(s,t,G \setminus \{v\})+4~.
\end{eqnarray*}
The claim follows.
\QED
%
For every new-ending path $P_{s,t,v}$, recall that $D^+_{s,t,v}=D_{s,t,v} \setminus \{b_{s,t,v}\}$. Let $(x,y)$ be the first missing in $P_{s,t,v} \setminus G_0$ (where $x$ is closer to $y$).
The following auxiliary claims are useful.
\begin{lemma}
\label{cl:clutone_suff}
For every vertex $u \in P_{s,t,v}$ such that $\LastE(P_{s,t,v}[s,u]) \notin G_0$, it holds that:
(a) $C_v(u)=C_1(u)$.
(b) $P_{s,t,v}[x, t] \subseteq D^+_{s,t,v}$.
\end{lemma}
\Proof
Begin with (a).
By the definition of the weight assignment $W$, it holds that $P_{s,t,v}[s,u]=P_{s,u,v}=SP(s,u, G \setminus \{v\},W)$. Since $\LastE(P_{s,t,v}[s,u]) \notin E^{local}$, it holds that $v \neq z_1(u)$, concluding that $z_v(u)=z_1(u)$ and hence $C_v(u)=C_1(u)$. Claim (a) follows. Consider claim (b). Let $b=b_{s,t,v}$. We show that $x \neq b$, which implies the claim. By \ref{cl:new}, $v \in \pi(s,y)$.
Since $b$ appears above $v$ on $\pi(s,t)$, $b$ is mutual to $\pi(s,t)$ and $\pi(s,y)$.
Hence, the $s-y$ shortest path has the following form: $\pi(s, y)=\pi(s,b) \circ \pi(b,v) \circ \pi(v,y)$.
Since $b \neq v \neq y$, $\dist(y, b,G)\geq 2$, and hence also $\dist(y, b,G\setminus\{v\})\geq 2$, concluding that $b \neq x$. The lemma follows.
\QED
\begin{corollary}
\label{cor:dplusnoi}
Let $t \in V_{\Delta}$. For every $P_{s,t,v} \in \mathcal{P}^{far}_{indep}(t)$, $P_{s,t,v}[x,t] \cap V_{f}(C_v(t))=\emptyset$ where $x$ is the first vertex of $D^+_{s,t,v}$.
\end{corollary}
\Proof
Since $P_{s,t,v}$ is independent, by Eq.
\ref{eq:depend}, $D^+_{\source,t,v} \cap T(t,\sourceset)=\{t\}$.
Since $t \in C_v(t)$, by Eq. (\ref{eq:hctau}),
$t \notin V_{f}(C_v(t))$ and hence $D^+_{\source,t,v} \cap V_{f}(C_v(t))=\emptyset$. The corollary follows by combining with Lemma \ref{cl:clutone_suff}(b).
\QED
\paragraph{Correctness analysis of $H_4(S)$.}
We now show that $H_4(\sourceset)$ is a $(4,\sourceset)$ FT-spanner.
\begin{lemma}
\label{lem:correct4sw}
$H_4(\sourceset)$ is a $(4,\sourceset)$ FT-spanner.
\end{lemma}
\Proof
Let $H=H_4(\sourceset)$.
The proof is by contradiction. Assume that there exists
a vertex $\source \in \sourceset$, a target vertex $t$
and a failing vertex $v$ such that $\dist(\source, t, H \setminus \{v\})> \dist(\source, t, G \setminus \{v\})+4$.
Let
\begin{eqnarray*}
BP=\{(t,v) \mid t \in V, v \in \left(V \cup \{\emptyset\}\right) \mbox{~and~}
\\
\dist(\source, t, H \setminus \{v\})> \dist(\source, t, G \setminus \{v\}+4\}
\end{eqnarray*}
be the set of ``bad pairs" namely, vertex pair $(t,v)$ whose additive stretch in $H$ is greater than 4 with respect to $\source\in \sourceset$. By the contradictory assumption $BP\ne \emptyset$.
Since the BFS tree $T_0(\source)$ is in $H$, it holds that the failing vertex $v \in \pi(\source,t)$ for every pair $(t,v) \in BP$.

For every bad pair $(t,v)\in BP$ define
$e_{t,v}$ to be the last missing edge of $P_{s,t,v}=SP(s,t, G \setminus \{v\},W)$ in $H$.
Let $d(t,v)=\dist(\source,e_{t,v},P_{s,t,v})$ be the distance of the last missing edge from $\source$ on  $P_{s,t,v}$.
Finally, let $(t_0,v_0) \in BP$ be the pair that minimizes $d(t,v)$, and let $ee_{t_0,v_0}=(u,t_i)$.
Note that $e_{t_0,v_0}$ is the {\em shallowest} ``deepest missing edge'' over all bad pairs $(t,v) \in BP$.
\begin{lemma}
\label{cl:isbad}
The pair $(t_i, v_0) \in BP$~.
\end{lemma}
\Proof
Assume towards contradiction that  $(t_i, v_0) \notin BP$ and let
$P'' \in SP(\source, t_i, H \setminus \{v_0\})$.
Hence, since  $(t_i, v_0) \notin BP$, it holds that
\begin{eqnarray}
\label{eq:up_cor_add}
|P''| &\leq& \dist(\source, t_i, G \setminus \{v_0\})+4
\\ &=& |P_{s,t_0, v_0}[\source, t_i]|+4. \nonumber
\end{eqnarray}
We now consider the following $\source-t_0$ replacement path
$Q=P'' \circ P_{s,t_0, v_0}[t_i,t_0]$.
By definition of $(t_i,v_0)$ (last missing edge on $P_{s,t_i,v_0}$), $Q \subseteq H \setminus \{v_0\}$. In addition,
\begin{eqnarray*}
|Q|&=&|P''|+|P_{t_0, v_0}[t_i,t_0]|
\leq
|P_{s,t_0, v_0}[\source, t_i]|+4+|P_{s,t_0, v_0}(t_i, t_0)|
\\&=&
|P_{s,t_0, v_0}|+4 ~=~ \dist(s, t_0, G \setminus \{v_0\})+4~,
\end{eqnarray*}
where the inequality follows by Eq. (\ref{eq:up_cor_add}).
This contradicts the fact that $(t_0,v_0) \in BP$.
\QED
Since $\pi(\source, t_i) \subseteq H$, by the fact that $(t_i,v_0) \in BP$, we have that the failing vertex $v_0$ occurs on the shortest-path $\pi(\source, t_i)$.
Since the last edge of $P_{\source,t_i,v_0}=P_{s,t_0,v_0}[s,t_i]$ is missing, by the fact that the clustering graph $G_{\Delta}$ is in $H$, by Obs. \ref{obs:clust_prop}(2), it holds that $t_i$ is clustered (i.e., $t_i \in V_{\Delta}$).
By step (2), since $E^{local} \subseteq H$, it holds that $v_0  \notin \{z_1(t_i), \LCA(s,C_1(t_i))\}$. Combining with Lemma \ref{lem:near}, it holds that $v_0 \notin V^{near}(s,t_i)$. Hence, $v_0 \in V^{far}(s,t_i)$.
There are further two cases.
Case (1) $P_{s,t_i,v_0}$ is dependant. By step (4.1), we then have that $\LastE(P_{s,t_i,v_0}) \in E^{far}_{dep}(t)$, contradiction to the fact that $\LastE(P_{s,t_i,v_0})$ is missing in $H$.

Consider case(2) where $P_{s,t_i,v_0}$ is an \emph{independent} path, i.e., $P_{\source, t_i, v_0} \in \mathcal{P}^{far}_{indep}(t)$ and hence it was considered to be bought in the path-buying procedure of Step (4.2).
By the fact that the pair $(t_i, v_0)$ is a bad pair (i.e., $(t_i, v_0) \in BP$), we conclude that the algorithm did not buy the path. Let $\tau$ be the iteration at which $P_{\tau}=P_{\source, t_i, v_0}$ was
considered to be purchased in the path-buying procedure.
Let $G_\tau$ be the current spanner in iteration $\tau$.
Let $x$ be the vertex incident to the first missing edge on $P_{\tau} \setminus E(G_{\tau})$.

Let $C_0=C_{v_0}(t_i)$ be the cluster of $t_i$ in $G_\Delta \setminus \{v_0\}$. Since $\LastE(P_{\tau}) \notin E^{local}$, by Lemma \ref{cl:clutone_suff}(a) $C_0=C_1(t_i)$ and $z_{v_0}(t_i)=z_1(t_i)$.
By definition,
$$v_0 \in \pi^{far}(s,t_i)=\pi(\source, C_0)\setminus \{\LCA(\source,C_0)\}.$$
Hence, in particular failing vertex is \emph{not} in the cluster $C_0$, i.e., $v_0 \notin C_0$ and by Eq. (\ref{eq:hctau}),
\begin{equation}
\label{eq:vin}
v_0 \in V_{f}(C_0).
\end{equation}

Since $P_{\tau}$ was not bought by the algorithm, by Eq. (\ref{eq:pathbuying6add}), we have that
\begin{equation}
\label{eq:cor4}
\dist(C_{v_0}(x), C_0, G_\tau \setminus V_{f}(C_0))\leq
\dist(x, t_i, P_\tau).
\end{equation}
Let $w_1 \in C_{v_0}(x)$ and $w_2 \in C_0$ be an arbitrary closest pair in $G_\tau \setminus V_{f}(C_0)$ from the clusters $C_{v_0}(x)$ and $C_0$ respectively satisfying that $\dist(w_1, w_2, G_\tau \setminus V_{f}(C_0))=\dist(C_{v_0}(x), C_0, G_\tau \setminus V_{f}(C_0))$.

\begin{figure}[htbp]
\begin{center}
\includegraphics[scale=0.4]{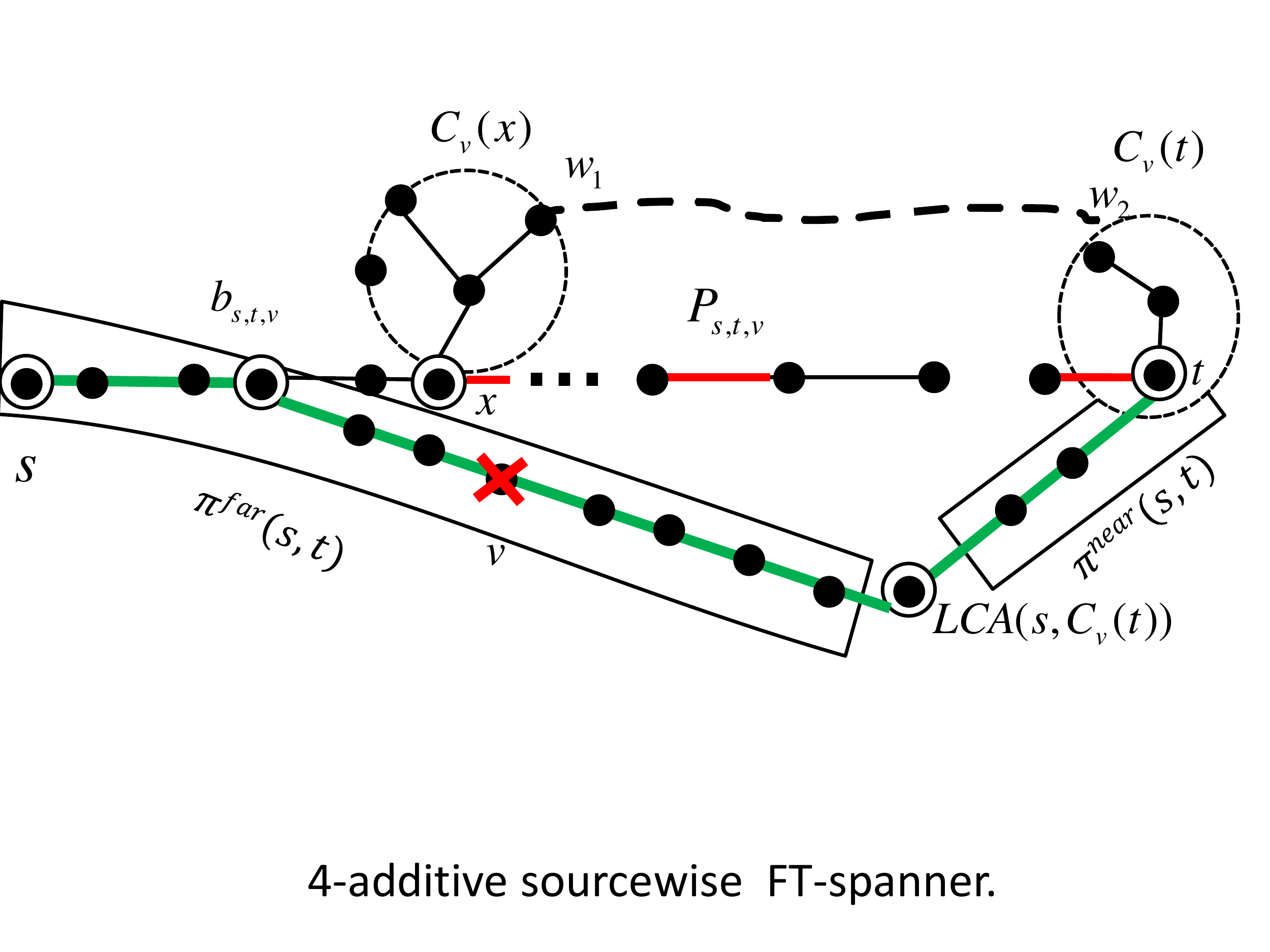}
\caption{Schematic illustration of the path-buying procedure of Alg.  \Conss. Shown is an $s-t$ path $P_{\tau}=P_{s,t,v}$ considered to be bought in time $\tau$. The green paths correspond to shortest-paths in $T_0(s)$ and the red edges correspond to missing edges on $P_{\tau} \setminus E(G_{\tau})$. The first missing edge on $P_{\tau} \setminus E(G_{\tau})$ is incident to $x$. If $P_{\tau}$ was not bought, then there exists a short route between a pair of vertices $w_1$ and $w_2$ belonging to $C_{v}(x)$ and $C_{v}(t)$ (respectively) in $H \setminus \{v\}$.
\label{fig:4add}}
\end{center}
\end{figure}

\begin{figure}[h!]
\begin{center}
\includegraphics[scale=0.25]{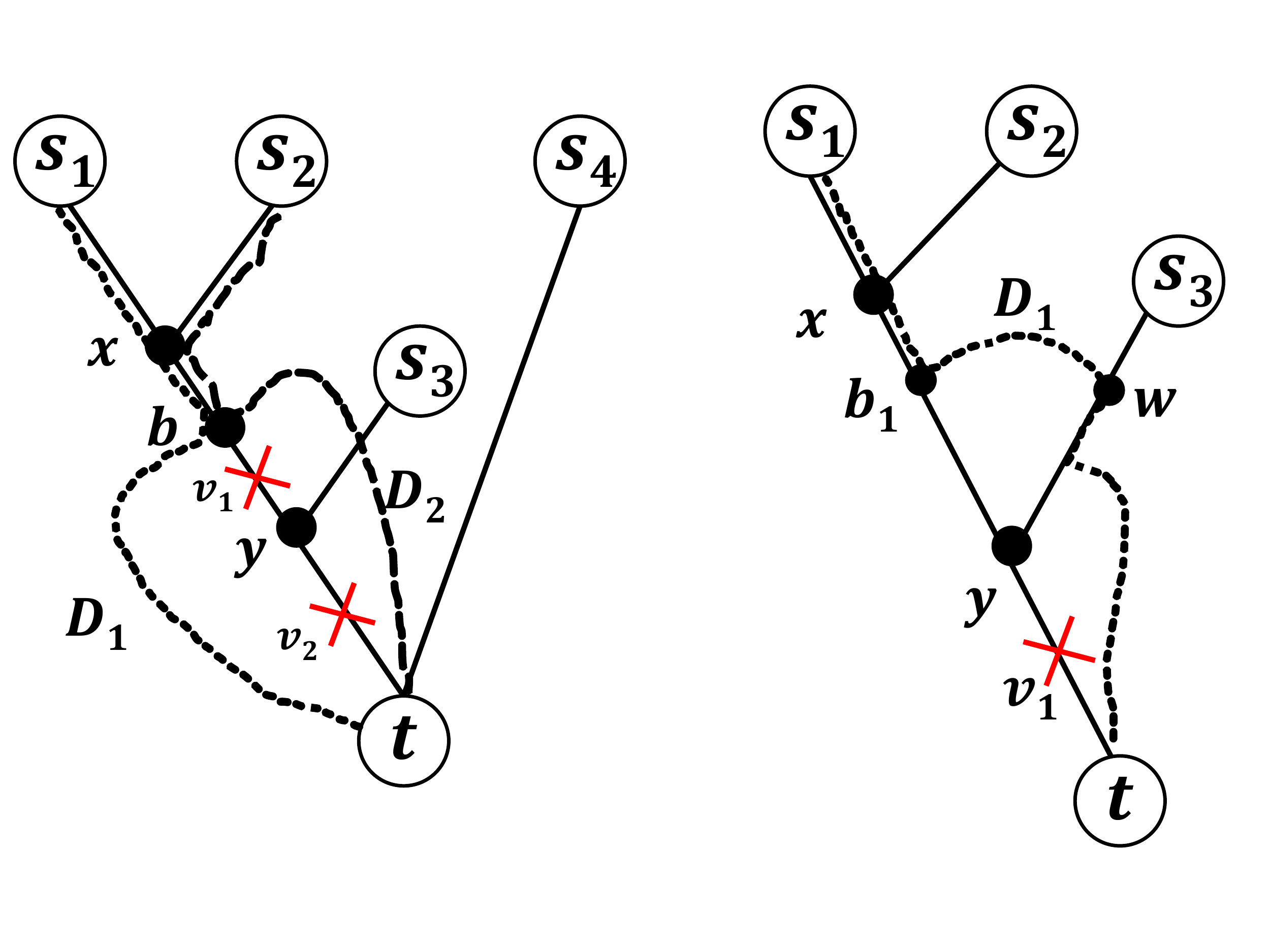}
\caption{Schematic illustration of new-ending dependent paths. Shown is the tree $T(t,S)$ with the root $t$ at the bottom and leaf set is contained in the set of sources $S$. (a) The two replacement paths have the same divergence point $b$, hence one of the new last edges is redundant.
(b) A new-ending $s_1-t$ dependant path $P_{s_1,t,v_1}$ with a divergence point $b_1 \in \pi(x,y)$ intersects with $\pi(s_3,t)$ at the vertex $w \notin \{b_1,t\}$.
Since $P_{s_1,t,v}$ is a new-ending path (i.e., its last edges is not on $T(t,S)$), the failing vertex $v$ must occur on the path $\pi(w,t)$. Hence $v_1\in \pi(s_1,t) \cap \pi(s_3,t)$, implying that $v_1\in \pi(y,t)$. Since this holds for any new-ending path with a divergence point in $\pi(x,y)$, we get that only one new edge from all these paths is needed.
\label{fig:xy}}
\end{center}
\end{figure}

Let $z_1$ (resp., $z_2$) be the cluster center of $C_{v_0}(x)$ (resp., $C_0$).
Consider the following $\source-t_i$ replacement path in $H\setminus \{v_0\}$, $P_5=P_1 \circ P_2 \circ P_3 \circ P_4$ where
$P_1=P_\tau[s, x]$, $P_2=[x, z_1, w_1]$ and
$P_3 \in SP(w_1, w_2, G_\tau \setminus V_{f}(C_0))$ and $P_4=[w_2, z_2, t_i]$. For an illustration see Fig.  \ref{fig:4add}.
We first claim that $P_5 \subseteq H \setminus \{v_0\}$. Since $x$ incident to the first missing edge on $P_{\tau}$, $P_1$ is in $H \setminus \{v_0\}$.
By Eq. (\ref{eq:vin}), $v_0 \in V_f(C_0)$ and since $w_1,w_2\subseteq G_\tau \setminus V_{f}(C_0)$ it also holds that $w_1,w_2 \neq v_0$. Finally note that $G_{\tau},G_{\Delta} \subseteq H$, hence $P_2,P_4 \subseteq H \setminus \{v_0\}$.
We next bound the length of $P_5$.
\begin{eqnarray}
\dist(\source, t_i, H \setminus \{v_0\})&\leq & P_5  ~\leq~
\dist(\source, x, G \setminus \{v_0\})+2+\dist(w_1, w_2, G_\tau \setminus V_{f}(C_0))+2 \nonumber
\\&=&
\dist(\source, x, G \setminus \{v_0\})+\dist(C_{v_0}(x), C_0, G_\tau \setminus V_{f}(C_0))+4 \nonumber
\\&\leq&
\dist(\source, x, G \setminus \{v_0\})+\dist(x, t_i, P_\tau)+4 \label{eq:add6sst2}
\\&=&
|P_{s,t_i,v}|+4=\dist(s, t_i, G \setminus \{v_0\})+4, \nonumber
\end{eqnarray}
where Eq. (\ref{eq:add6sst2}) follows by Eq. (\ref{eq:cor4}). We end with contradiction to the fact that the pair $(t_i, v_0)$ is a bad pair. The claim follows.
\QED
\paragraph{Size analysis of $H_4(S)$.}
We proceed with the size analysis.
\begin{lemma}
\label{fc:nearedge}
For every $t \in V_{\Delta}$, $|E^{local}(t)|=O(|\sourceset|)$, hence $|E^{local}|=O(|\sourceset| \cdot n)$.
\end{lemma}
\paragraph{Bounding the number of last edges in $E^{far}_{dep}(t)$.}
We now turn to bound the number of edges added due to step (4.1), i.e., the last edges of new-ending \emph{dependent} paths $P_{\source,t,v}$ protecting against the faults in the far segment $\pi^{far}(\source,t)$.
To bound the number of edges in $E^{far}_{dep}(t)$, consider the partial BFS tree rooted at $t$, $T(t,\sourceset)\subseteq T_0(T)$, whose leaf set is contained in the vertex set $\sourceset$ where $T(t,\sourceset)=\bigcup_{\source\in \sourceset} \pi(\source,t)$. It is convenient to view this tree as going from the leafs towards the root, where the root $t$ is at the bottom and the leafs are on the top of the tree. Let $V^+=\sourceset \cup \{u \in T(t,\sourceset) ~\mid~ \deg(u, T(t,\sourceset))\geq 3\},$
be the union of $\sourceset$ and the vertices with degree at least 3 in the tree $T(t,\sourceset)$. We have that $|V^+| <2|\sourceset|$.
%
A pair of vertices $x, y \in  V^+$ is \emph{adjacent} if their shortest-path $\pi(x,y)$ is contained in the tree $T(t,\sourceset)$ and it is free from any other $V^+$ vertex, i.e,
$\pi(x,y) \subseteq T(t,\sourceset)$ and $\pi(x,y) \cap V^+=\{x,y\}$.
Let $\Pi(V^+)=\{\pi(x,y) ~\mid~ x, y \in V^+ \mbox{~and~} x, y \mbox{~are adjacent~} \}$ be the collection of paths between adjacent pairs.
\begin{observation}
\label{obs:xyt}
(1) $T(t,\sourceset)=\Pi(V^+)$.
(2) $\Pi(V^+)$ consists of at most $2|\sourceset|+1$ paths $\pi(x,y)$ (i.e., there are at most $2|\sourceset|$ adjacent pairs).
\end{observation}
We now show the following.
\begin{lemma}
\label{lem:depn_bound4}
For every $t \in V_{\Delta}$, $|E^{far}_{dep}(t)|=O(|\sourceset|)$.
\end{lemma}
%
We first claim that every two dependent replacement paths with the same divergence point have the same last edge. 
\begin{lemma}
\label{cl:divergence_local}
For every two dependent paths $P_{\source_1,t, v_1}, P_{\source_2,t, v_2} \in \mathcal{P}^{far}_{dep}(t)$, if $b_{\source_1,t, v_1} = b_{\source_2,t, v_2}$ then $\LastE(P_{\source_1,t, v_1}) =\LastE(P_{\source_1,t, v_2})$.
\end{lemma}
\Proof
Let $b=b_{s_1,t,v_1}=b_{s_2,t,v_2}$.
Since $b \in \pi(s_1,t) \cap \pi(s_2,t)$ it holds that $\pi(s_i,t)=\pi(s_i, b) \circ \pi(b,t)$ for $i \in \{1,2\}$. In addition, since $P_{\source_i,t, v_i}[s_i,b]=\pi(s_i,b)$ for $i \in \{1,2\}$, it holds that both failing vertices $v_1$ and $v_2$ occur in the common segment $\pi(b,t)$. Recall that $P_{\source_i,t, v_i}$ is a new-ending path, hence by the definition of the divergence point $b$ (see Lemma \ref{cl:new}(b)), it holds that $V(P_{s_i,t,v_i}[b,t]) \cap V(\pi(b,t))=\{b,t\}$ and hence both detours are free from the failing vertices.
Hence, $P_{\source_1,t, v_1}[b,t], P_{\source_1,t, v_1}[b,t]=SP(b,t, G \setminus\{v_1,v_2\})$.
We get that $\LastE(P_{\source_1,t, v_1}) =\LastE(P_{\source_2,t, v_2})$ as needed.
\QED
Since our goal is to bound the number of last edges of the new ending dependent paths $\mathcal{P}^{far}_{dep}(t)$, to avoid double counting, we now restrict attention to $\mathcal{Q}^{far}(t)$, a collection of representative paths in $\mathcal{P}^{far}_{dep}(t)$ each ending with a distinct new edge from $E^{far}_{dep}(t)$. Formally,
for each new edge $e \in E^{far}_{dep}(t)$, let $P(e)$ be an arbitrary path in $\mathcal{P}^{far}_{dep}(t)$ satisfying that $\LastE(P(e))=e$.
Let $\mathcal{Q}^{far}(t)=\{P(e), e \in E^{far}_{dep}(t)\}$ (hence $|\mathcal{Q}^{far}(t)|=|E^{far}_{dep}(t)|$). From now on, we aim towards bounding the cardinality of  $\mathcal{Q}^{far}(t)$.
Let $\DP=\{b_{\source,t,v} ~\mid~ P_{\source,t,v} \in \mathcal{Q}^{far}(t)\}$
be the set of divergence points of the new ending paths in $\mathcal{Q}^{far}(t)$.
By Lemma \ref{cl:divergence_local}, it holds that in order to bound the cardinality of $\mathcal{P}^{far}_{dep}(t)$, it is sufficient to bound the number of distinct divergence points.
To do that, we show that every path $\pi(x,y)$ of two adjacent vertices $x,y \in V^+$, contains at most one divergence point in $\DP \setminus V^+$.
\begin{lemma}
\label{cl:adj_one}
$|\pi(x,y) \cap (\DP \setminus V^+)| \leq 1$ for every $\pi(x,y) \in \Pi(V^+)$.
\end{lemma}
\Proof
Assume, towards contradiction, that there are two divergence points $b_{\source_1,t,v_1}$ and $b_{\source_2,t,v_2}$ on some path $\pi(x,y)$ for two adjacent vertices $x,y \in V^+$. For ease of notation, let $P_i=P_{\source_i,t,v_i}, b_i=b_{\source_i,t,v_i}$, $D_i=D_{\source_i,t,v_i}$ and $D^+_i=D_i \setminus \{b_i\}$ for $i \in \{1,2\}$.
Without loss of generality, assume the following: (1) $y$ is closer to $t$ than $x$ and (2) $b_{2}$ is closer to $t$ than $b_{1}$.  By construction, the vertices $s_1$ and $s_2$ are in the subtree $T(x) \subseteq T(t,\sourceset)$.
For an illustration see Fig. \ref{fig:xy}.
We now claim that the failing vertices $v_1,v_2$ occur on $\pi(y, t)$. Since $D^+_1$ and $D^+_2$ are vertex disjoint with
$\pi(y,t)\setminus \{t\}$, it would imply that both detour segments $D_1$ and $D_2$ are free from the failing vertices and hence at least one of the two new edges $\LastE(P_1), \LastE(P_2)$ could have been avoided.
We now focus on $v_1$ and show that $v_1 \in \pi(y,t)$, the exact same argumentation holds for $v_2$.
Since $P_{1}$ is a new-ending \emph{dependent} path, by Eq. (\ref{eq:depend}), there exists some source $\source_3 \in \source \setminus \{\source_1\}$ satisfying that $\left(D^+_1 \cap \pi(\source_3, t)\right)\setminus\{t\} \neq \emptyset$.
Let $w \in \left(D^+_1 \cap \pi(\source_3, t)\right)\setminus\{t\}$ be the first intersection point (closest to $\source_1$). See Fig. \ref{fig:xy} for schematic illustration.
We first claim that $\source_3$ is not in the subtree $T(x) \subseteq T(t,\sourceset)$ rooted at $x$.
To see why this holds, assume, towards contradiction, that $\source_3 \in T(x)$. It then holds that the replacement path $P_1$ has the following form
$P_1=\pi[\source_1, x] \circ \pi(x, b_{1}) \circ P_1[b_{1}, w] \circ P_1[w, t]$. Recall, that since $b_{1} \in \DP \setminus V^+$, $b_{1} \neq x$ and also $b_1\neq w$. Since $P_1[x,w]$ goes through $b_1$, by the optimality of $P_1$, it holds that
\begin{equation}
\label{eq:n1q}
\dist(x,w,G \setminus \{v_1\})>\dist(b_1,w,G \setminus \{v_1\})~.
\end{equation}
On the other hand, the path $\pi(\source_3, t)$ has the following form: $\pi(\source_3,t)=\pi(\source_3,w) \circ \pi(w,x) \circ \pi(x,b_1)\circ \pi(b_1,t)$. Hence, $\pi(w,b_1)$ goes through $x$. Since the failing vertex $v_1 \in \pi(b_1,t)$ is not in $\pi(w,b_1)$, by the optimality of $\pi(w,b_1)$, we get that
$\dist(w,b_1,G \setminus \{v_1\})>\dist(x,w,G \setminus \{v_1\})$, leading to contradiction with Ineq. (\ref{eq:n1q}).
%
Hence, we conclude that $\source_3 \notin T(x)$ (in particular this implies that $\source_3 \neq \source_2$).
Note that $\pi(w,t)$ is a segment of $\pi(\source_3,t)$ and hence it is contained in the tree $T(t,\sourceset)$. Since $P_1$ is a new-ending path (i.e., $\LastE(P_1) \notin T(t,\sourceset)$), we have that $P_{1}[w,t]\neq \pi(w,t)$ are distinct $w-t$ paths.
We next claim that the failing vertex $v_1$ must occur on $\pi(w,t)$ and hence also on $\pi(\source_3,t)$. To see this, observe that
if $\pi(w,t)$ would have been free from the failing vertex $v_1$, then it implies that $\pi(w,t)=SP(w,t, G \setminus \{v_1\})=P_1[w,t]$, contradiction as $\LastE(P_1) \neq \LastE(\pi(w,t))$.
Finally, we show that $v_1 \in \pi(y,t)$.
By the above, the failing vertex $v_1$ is common to both paths $\pi(\source_1,t)$ and $\pi(\source_3,t)$, i.e., $v_1 \in \pi(\source_1,t) \cap \pi(\source_3,t)$. By the definition of the path $\pi(x,y)$, all its internal vertices $u$ have degree $2$ and hence $(\pi(x,y) \cap \pi(\source_3,t)) \setminus \{y\}=\emptyset$, concluding that $v_1 \in \pi(y, t)$. By the same argumentation, it also holds that $v_2$ is in $\pi(y,t)$.
%
As the detours $D_{1}$ and $D_{2}$ are vertex disjoint with $\pi(y,t) \setminus \{t\}$, it holds that they are free from the two failing vertices, i.e., $v_1,v_2 \notin D_{1} \cup D_{2}$.
Since $P_1,P_2 \in \mathcal{Q}^{far}(t)$, it holds that
$\LastE(P_{1}) \neq \LastE(P_{2})$, and hence there are two
$b_{1}-t$ distinct shortest paths in $G \setminus \{v_1, v_2\}$, given by $D_{1}$ and $\pi(b_1,b_2) \circ D_{2}$. By optimality of these paths, they are of the same lengths. Again, we end with contradiction to the uniqueness of the weight assignment $W$.
The claim follows.
\QED
By Lemma \ref{cl:divergence_local} there are at most $|V^+|$ replacement paths with divergence point in $V^+$. By Lemma \ref{cl:adj_one}, there is at most one divergence point on each segment $\pi(x,y)$ of an adjacent pair $(x,y)$. Combining with Obs. \ref{obs:xyt}(2), we get $|E^{far}(t)|=|\mathcal{Q}^{far}(t)|=O(|\sourceset|)$. The lemma follows.
\QED
We complete the size analysis and proves Lemma \ref{lem:h4}, by bounding the number of edges added by the path-buying procedure of Step (4.2).
\paragraph{Bounding the number of edges added due to the path-buying procedure.}
Finally, it remains to bound the number of edges added due to the path-buying procedure of step (4.2).
\begin{lemma}
\label{lem:hsize6}
$|H_4(\sourceset)\setminus G_0|=\widetilde{O}((n/|S|)^3)$.
\end{lemma}
\Proof
Let $\mathcal{B} \subseteq \mathcal{P}^{far}_{indep}(t)$ be the set of paths bought in the path-buying procedure of Step (4.2).
For every \emph{ordered} pair of clusters $C_1, C_2 \in \mathcal{C}$, let $\mathcal{B}(C_1, C_2) \subseteq \mathcal{B}$ be the set of paths that were added since they improved the distance of $C_1$ and $C_2$, that is
$$\mathcal{B}(C_1, C_2)=\{ P_{\tau} \in \mathcal{B} ~\mid~ C_1(\tau)=C_1 \mbox{~and~} C_2(\tau)=C_2\}$$
Clearly, $\mathcal{B}=\bigcup_{C_1, C_2 \in \mathcal{C}} \mathcal{B}(C_1, C_2)$.
We next use the fact that the diameter of each cluster $C \in \mathcal{C}$ is small, to bound the cardinality of the set $\mathcal{B}(C_1, C_2)$.
\begin{lemma}
$|\mathcal{B}(C_1, C_2)| \leq 5$ for every $C_1, C_2 \in \mathcal{C}$.
\end{lemma}
\Proof
Fix $C_1, C_2 \in \mathcal{C}$ and order the paths of $\mathcal{B}(C_1, C_2)$ according to the time step they were added to the spanner $\mathcal{B}(C_1, C_2)=\{P_{\tau_1}, \ldots, P_{\tau_N}\}$ where $\tau_1 < \tau_2 < \ldots < \tau_N$ where $N=|\mathcal{B}(C_1, C_2)|$.
Since $P_{\tau_k} \in \mathcal{P}^{far}_{indep}$, it is a new-ending path, i.e., $\LastE(P_{\tau_k}) \notin T_0(\sourceset)$.
Let $P_{\tau_k}=P_{\source_k,t_k,v_k}$ and $D_{\tau_k}=D_{\source_k,t_k,v_k}$ denote the detour segment of this path. Let $x_k$ be the vertex adjacent to the first missing edge on $P_{\tau_k}$. Hence, $C_1=C_{v_k}(x_k)$ and $C_2=C_{v_k}(t_k)$ for every $k \in \{1, \ldots, N\}$ and also $V_{f}(C_{v_k}(t_k))=V_{f}(C_2)$ for every $k \in \{1,\ldots, N\}$.
Since $T_0(\sourceset) \subseteq G_0$, the missing edges of $P_{\tau_k}$ are restricted to the detour segment $D_{\tau_k}$.

In addition, since $P_{\tau_k} \in \mathcal{P}^{far}_{indep}$, it holds that the failing vertex $v_k$ occurs on the far segment $\pi^{far}(s_k,t_k)$ and in particular, $v_k \notin C_2$ (i.e., $v_k$ occurs strictly above the least common ancestor $\LCA(s_k, C_2)$ and since all cluster members appears on $T_0(s_k)$ in the subtree rooted at $\LCA(s_k,C_2)$, the far segment $\pi^{far}(s_k,t_k)$ is free from cluster members).
We therefore have
\begin{equation}
\label{eq:pb_failedv}
\{v_1, \ldots, v_N\}\subseteq  V_f(C_2)~.
\end{equation}
Note that $\pi(\source_k, C_2) \subseteq \pi(\source_k,t_k)$ and hence it is contained in $T(t_k,\sourceset)$ for every $k \in \{1,\ldots,N\}$. By the definition of independent paths (see Eq. (\ref{eq:depend}) for the definition of \emph{dependent} paths), we have that
\begin{equation}
\label{eq:nottoachs}
D^+_{\source_k,t_k,v_k} \cap T(t_k,\sourceset)=\{t_k\}~.
\end{equation}
Consequently, by Lemma \ref{cl:clutone_suff}(b),
\begin{equation}
\label{eq:pb_notinhc}
P_{\tau_k}[x_k,t_k] \subseteq D^+_{\tau_k} \subseteq G \setminus V_{f}(C_2)~.
\end{equation}
where the last inclusion holds by the fact that $t_k \notin V_{f}(C_2)$ and
$V_{f}(C_2) \subseteq \bigcup_{s \in S} \pi(s, C_2)
\subseteq \bigcup_{s \in S} \pi(s, t)=T(t,S)$.
Let $z_i \in Z$ be the cluster center of $C_i$ for $i \in \{1,2\}$.
We therefore have that $z_1=z_{v_1}(x_1) =\ldots =z_{v_N}(x_N)$  and $z_2=z_{v_1}(t_1)=\ldots =z_{v_N}(t_N)$ and hence $z_2 \neq v_k$ for every $k \in \{1, \ldots, N\}$.
Hence,
\begin{equation}
\label{eq:clustercenternotfull}
z_1,z_2 \notin \{v_1, \ldots, v_N\}.
\end{equation}
For every $k \in \{1,\ldots, N\}$, denote
$$X_k=\dist(x_k, t_k, G_{\tau_{k+1}} \setminus V_{f}(C_2)).$$
We now show that $X_k<X_{k-1}$ for every $k \in \{2, \ldots N\}$.

Since the path $P_{\tau_k}$ is purchased at time $\tau_k$, we have that
\begin{eqnarray}
X_k &\leq&
\dist(x_k, t_k, P_{\tau_k} \setminus V_{f}(C_2)) \label{eq:adub_1}
\\&=&
\dist(x_k, t_k, P_{\tau_k}) \label{eq:adub_2}
\\&<&
\dist(C_{1}, C_{2}, G_{\tau_{k}} \setminus V_{f}(C_2)) \label{eq:adub_3}
\\&\leq&
X_{k-1}~,
\label{eq:adub_4}
\end{eqnarray}
where Eq. (\ref{eq:adub_1}) follows by the fact that $P_{\tau_k} \subseteq G_{\tau_{k+1}}$, Eq. (\ref{eq:adub_2}) follows by Eq. (\ref{eq:pb_notinhc}). Eq. (\ref{eq:adub_3}) follows by the fact that $P_{\tau_k}$ was bought and by Eq. (\ref{eq:pathbuying6add}), and Eq. (\ref{eq:adub_4}) follows by the fact that $x_{k-1} \in C_1$ and $t_{k-1} \in C_2$.

Therefore, 
we have that
\begin{equation}
\label{eq:up_srtect1}
X_N \leq  X_1-(N-1)~.
\end{equation}
Conversely, we have that
\begin{eqnarray}
X_N
&\geq&
\dist(x_{N}, t_N, G \setminus \{v_1, \ldots, v_N\}) \label{eq:up_t01}
\\
&\geq&
\dist(x_1, t_1, G \setminus \{v_1, \ldots, v_N\})-4 \label{eq:up_t02}
\\
& =&
\dist(x_1, t_1, P_{\tau_1})-4 \label{eq:up_t03}
= \dist(x_1, t_1, P_{\tau_1} \setminus V_{f}(C_2))-4
\\& \geq &
X_1-4~, \label{eq:up_t04}
\end{eqnarray}
where Eq. (\ref{eq:up_t01}) follows as $G_{\tau_{N+1}} \subseteq G$ and by Eq. (\ref{eq:pb_failedv}), $\{v_1, \ldots, v_N\} \subseteq V_{f}(C_2)$.
To see Eq. (\ref{eq:up_t02}), we need to prove the existence of the intracluster paths $R_1=[x_1, z_1, x_N]$ and $R_2=[t_1, z_2, t_N]$ in $G \setminus \{v_1, \ldots, v_N\}$ where $z_1$ (resp., $z_2$) is the cluster center of $C_1$ (resp., $C_2$).
By definition, $x_{1},x_N \in C_{1}$ and $t_1,t_N \in C_{2}$. Hence, $z_1$ (resp., $z_2$) is a common neighbor of both $x_{1}$ and $x_N$ (resp., $t_1$ and $t_N$).
By (\ref{eq:clustercenternotfull}),
$z_1,z_2 \notin \{v_1, \ldots, v_N\}$.

In addition, by Eq. (\ref{eq:pb_notinhc}),
$x_k, t_k \in G \setminus V_{f}(C_2)$ for every $k\in \{1, \ldots, N\}$ and by Eq. (\ref{eq:pb_failedv}),
it also holds that  $x_k,t_k \notin \{v_1, \ldots, v_N\}$ for every $k \in \{1,\ldots,N\}$. Hence, $R_1$ and $R_2$ exists in
$G\setminus \{v_{1},\ldots,v_N\}$ and Eq. (\ref{eq:up_t02}) follows by the triangle inequality.
Eq. (\ref{eq:up_t03}) follows by
Eq. (\ref{eq:pb_notinhc}) and Eq. (\ref{eq:pb_failedv}).
%
%
Finally, Eq. (\ref{eq:up_t04}) follows by the fact that $P_{\tau_1}$ was added at step $\tau_1$, hence $P_{\tau_1} \subseteq G_{\tau_{2}}$. We get that $N \leq 5$. The claim follows.
\QED
By Obs. \ref{obs:clust_prop}(2) and Obs. \ref{obs:nmissing},
every path $P_{\tau_k}$ contains at most $O(n/|\Delta|)=O(n/|S|)$ missing edges in $G \setminus G_{\Delta}$.
Hence,
\begin{eqnarray}
\label{eq:h6size}
|E(G_{\tau'} \setminus G_0)|&=&O(n/|S|) \cdot |\mathcal{B}|
\\&=&
O(n/|S|) \cdot \sum_{C_1, C_2 \in \mathcal{C}} |\mathcal{B}(C_1, C_2)|
\\& \leq &
O(n/|S|)\cdot |\mathcal{C}|^2=\widetilde{O}((n/|S|)^3)~.
\end{eqnarray}
where the last equality follows by the fact that $|\mathcal{C}|=\widetilde{O}(n/|S|)$. The claim follows.
\QED
%
\subsection{Sourcewise spanner with additive stretch 8}
In this section, we present Alg. \Consssix\ for constructing a sourcewise additive FT-spanner with additive stretch $8$. The size of the resulting spanner is smaller (in order) than the $H_4(S)$ spanner of Alg. \Conss, at the expense of larger stretch.
The algorithm is similar in spirit to Alg. \Conss\ and the major distinction is in the path-buying procedure of step (4.2).
\begin{lemma}
\label{lem:h8}
There exists a subgraph $H_{8}(\sourceset) \subseteq G$ with $\widetilde{O}(\max\{|\sourceset| \cdot n, (n/|\sourceset|)^{2}\})$ edges
s.t.
$\dist(\source,t, H_{8}(\sourceset) \setminus \{v\})\leq \dist(\source,t, G \setminus \{v\})+8$ for every
$(s, t) \in S \times V$ and every $v \in V$.
\end{lemma}
\subsubsection{Algorithm \Consssix\ for constructing $H_8(\sourceset)$ spanner}
\paragraph{Step (0-4.1):} Same as in Alg. \Conss.
Let $E^{local}, E^{far}_{dep}$ be the set of last edges obtained at the end of step (3) and set (4.1) respectively.
Let $\mathcal{P}^{far}_{indep}$ be the set of
new-ending independent paths.
\paragraph{Step (4.2): Handling \emph{independent} new-ending paths.}
Starting with $G_0$ as in Eq. (\ref{eq:g0}), the paths of $\mathcal{P}^{far}_{indep}$ are considered in an arbitrary order.
At step $\tau$, we are given $G_\tau \subseteq G$ and consider the path $P_{\tau}=P_{s,t,v}$. Let $D_\tau=P_\tau\setminus \pi(s,t)$ be the detour segment of $P_{\tau}$ (since $\pi(s,t) \subseteq T_0(S)$ is in $G_0$, all missing edges of $P_{\tau}$ occur on its detour segment).
\par To decide whether $P_{\tau}$ should be added to $G_\tau$, the number of pairwise cluster ``distance improvements" is compared to the number of new edges added due to $P_{\tau}$. To do that we compute the set $\ValSet(P_{\tau})$ containing all pairs of clusters that achieves a better distance if $P_{\tau}$ is bought.
The value and cost of $P_{\tau}$ are computed as follows.
Let $\Value(P_{\tau})=|\ValSet(P_{\tau})|$ as the number of distance improvements as formally defined later. We next define a key vertex $\phi_\tau \in V_{\Delta}$ on the path $P_{\tau}$.
\begin{definition}
\label{def:phi}
Let $\phi_{s,t,v}$ (or  $\phi_{\tau}$  for short) be the last vertex on $P_{\tau}$ (closest to $t$) satisfying that: (N1) $\LastE(P_{\tau}[s, \phi_{\tau}]) \notin G_{\tau}$, and (N2)
$v \in \pi^{near}(s,\phi_\tau)=\pi(\ell, \phi_{\tau})\setminus \{\ell\}$ where $\ell=\LCA(s, C_v(\phi_{\tau}))$.\\
If there is no vertex on $P_{\tau}$ that satisfies both (N1) and (N2), then let $\phi_{\tau}$ be the first vertex incident to the first missing edge on $P_{\tau} \setminus E(G_\tau)$ (i.e., such that $P_{\tau}[s,\phi_{\tau}]$ is the maximal prefix that is contained in $G_{\tau}$).
\end{definition}
Let $Q_{\tau}=P_{\tau}[\phi_{\tau},t]$ and define $\Cost(P_\tau)=|E(Q_{\tau}) \setminus E(G_\tau)|$ be the number of edges of $Q_{\tau}$ that are missing in the current subgraph $G_\tau$. Thus $\Cost(P_\tau)$ represents the increase in the size of the spanner $G_\tau$ if the procedure adds $Q_\tau$. Our algorithm attempts to buy only the suffix $Q_\tau$ of $P_{\tau}$ when considering $P_{\tau}$.
We now define the set $\ValSet(P_{\tau}) \subseteq \mathcal{C} \times \mathcal{C}$ which contains a collection \emph{ordered} cluster pairs.
Let $C_1(\tau)=C_v(\phi_{\tau})$ and $C_2(\tau)=C_v(t)$ be the clusters of $\phi_{\tau}$ and $t$ in $G_{\Delta} \setminus \{v\}$. Let $\kappa=\Cost(P_{\tau})$. The candidate $P_{\tau}$ is said to be \emph{cheap} if $\kappa\leq 4$, otherwise it is \emph{costly}. The definition of $\ValSet(P_{\tau})$ depends on whether or not the path is cheap.
In particular, if $P_{\tau}$ is cheap, then
let $\ValSet(P_{\tau})=\{(C_1(\tau), C_2(\tau))\}$ only if
\begin{equation}
\label{eq:short_pb}
\dist(\phi_{\tau}, t, P_\tau)< \dist(C_1(\tau), C_2(\tau), G_\tau \setminus V_{f}(C_2(\tau)))~,
\end{equation}
where $V_{f}(C_2(\tau))$ is as given by Eq. (\ref{eq:hctau}), and let $\ValSet(P_{\tau})=\emptyset$ otherwise.
Alternatively, if $P_{\tau}$ is costly, we do the following.

\begin{definition}
\label{def:u}
Let $U_{s,t,v}=\{u_{3\ell+1} ~\mid~ \ell \in \{0, \ldots, \lfloor (\kappa-1)/3 \rfloor\}\} \subseteq  Q_{\tau}$
be some representative endpoints of missing edges on $Q_{\tau}$ satisfying that
\begin{eqnarray}
\label{eq:z_up_add}
\LastE(Q_{\tau}[\phi_{\tau}, u_{\ell}]) \notin G_{\tau} \mbox{~for every~} \nonumber u_{\ell} \in U_{s,t,v}
\mbox{~and~}
\dist(u_{\ell}, u_{\ell'}, Q_{\tau})&\geq& 3~
\end{eqnarray}
for every $u_\ell, u_{\ell'} \in U_{s,t,v}$.
\end{definition}
Define
\begin{eqnarray}
\label{eq:contvs1}
\ValSet_1(P_{\tau})~=~\{(C_1(\tau), C_\ell) ~\mid~ C_\ell=C_{v}(u_\ell), u_{\ell} \in U_{s,t,v} \\\mbox{and}~
\dist(\phi_\tau, u_\ell, P_{\tau}) <
\dist(C_1(\tau), C_\ell, G_{\tau} \setminus V_{f}(C_\ell))\}\nonumber
\end{eqnarray}
and
\begin{eqnarray}
\label{eq:contvs2}
\ValSet_2(P_{\tau})~=~\{(C_\ell,C_2(\tau)) ~\mid~ C_\ell=C_{v}(u_\ell), u_{\ell} \in U_{s,t,v} \\\mbox{and}~
\dist(u_\ell, t, P_{\tau}) <
\dist(C_\ell, C_2(\tau), G_{\tau} \setminus V_{f}(C_2(\tau)))\}\nonumber
\end{eqnarray}
Let $\ValSet(P_{\tau})=\ValSet_1(P_{\tau})\cup \ValSet_2(P_{\tau})$.
The subpath $Q_\tau$ is added to $G_{\tau}$ resulting in $G_{\tau+1}$ only if
\begin{equation}
\label{eq:cost_value}
\Cost(P_\tau) \leq 4 \cdot \Value(P_\tau)~,
\end{equation}
where $\Value(P_\tau)=|\ValSet(P_{\tau})|$.
(Note that when $P_{\tau}$ is cheap, Eq. (\ref{eq:cost_value}) holds iff Eq. (\ref{eq:short_pb}) holds.)
The output of Alg. \Consssix\ is the subgraph
$H_8(S)=G_{\tau'}$ where $\tau'=|\mathcal{P}^{far}_{indep}|.$
This completes the description of the algorithm.

\paragraph{Analysis.}
Throughout the discussion, a path $P_{s,t,v}$ is a new-ending path, if $\LastE(P_{s,t,v})\notin G_0$ (see Eq. (\ref{eq:g0})). Hence, we consider only $P_{s,t,v} \in \mathcal{P}^{far}_{indep}(t)$ paths for clustered vertices $t \in V_{\Delta}$.

For a new-ending path $P_{s,t,v}$, recall that $b_{s,t,v}$ is the unique divergence point of $P_{s,t,v}$ and $\pi(s,t)$ and let $D_{s,t,v}$ be the detour segment, i.e., $D_{s,t,v}=P_{s,t,v}[b_{s,t,v},t]$ and $D^+_{s,t,v}=D_{s,t,v}\setminus \{b_{s,t,v}\}$.
Let $Q_{s,t,v}=P_{s,t,v}[\phi_{s,t,v}, t]$ be the path segment that was considered to be bought in step (4.2) (see Def. \ref{def:phi}).
\begin{observation}
\label{obs:dphi}
$Q_{s,t,v} \subseteq D^+_{s,t,v}$.
\end{observation}
\Proof
Let $x$ be the first vertex incident to a new-edge on $P_{s,t,v}$ (such that $P_{s,t,v}[s,x]$ is the maximal prefix that is contained in $G_0$). Since $\phi_{s,t,v}$ occurs not before $x$ on $P_{s,t,v}$ the observation follows by Lemma \ref{cl:clutone_suff}(b).
\QED

\begin{lemma}
\label{lem:rep_new_hc}
Let $P_{s,t,v} \in \mathcal{P}^{far}_{indep}(t)$ be a new-ending replacement path.
Then for every $u_k \in U_{s,t,v} \cup \{t\}$ with $C_k=C_v(u_k)$ it holds that:\\
(a) $C_k=C_1(u_k)$. \\
(b) $V(P_{s,t_i,v}[b_{s,t,v}, u_k]) \cap V(T(u_k,\sourceset))=\{b_{s,t,v},u_k\}$. \\
(c) $Q_{s,t,v}[\phi_{s,t,v},u_k]\cap V_{f}(C_k)=\{\emptyset\}$.\\
(d) $v \in V_{f}(C_k)$.
\end{lemma}
\Proof
We begin with (a).
By the uniqueness of the weight assignment $W$, $P_{s,t_i,v}[s,u_k]=P_{s,u_k,v}=SP(s, u_k, G \setminus \{v\},W)$. By the uniqueness of the divergence point $b_{s,t,v}$ and in particular by Lemma \ref{cl:new}(b),
\begin{equation}
\label{eq:bsame}
b_{s,t,v}=b_{s,u_k,v}~.
\end{equation}
Since $\LastE(P_{s,u_k,v}) \notin E^{local}$, concluding that $v \neq z_1(u_k)$ and hence $z_v(u_k)=z_1(u_k)$ and (a) holds.

Consider (b). By the definition of the set $U_{s,t,v}$ (see Def. \ref{def:u}), it holds $\LastE(P_{s,u_k,v}) \notin G_0$. Since $u_k \in Q_{s,t,v}$ occurs strictly after $\phi_{s,t,v}$, by the definition Def. \ref{def:phi}, it holds that $u_k$ did not satisfy property (N2). Hence,
since $\LastE(P_{s,u_k,v}) \notin E^{local}$, $v \notin \{z_1(u_k), \LCA(s, C_1(u_k))\}$ and hence
$v \in \pi^{far}(s,u_k)$. As $\LastE(P_{s,u_k,v}) \notin E^{far}_{dep}(u_k)$, we get that $P_{s,u_k,v}$ is a new-ending \emph{independent} path.
By Eq. (\ref{eq:depend}),
$V(P_{s,u_k,v}[b_{s,u_k,v}, u_k]) \cap V(T(u_k,\sourceset))=\{b_{s,u_k,v}, u_k\}$.
Hence (b) holds by Eq. (\ref{eq:bsame}).

\par We now turn to consider claim (c).
By Eq. (\ref{eq:hctau}), $V_{f}(C_v(u_k)) \subseteq T(u_k,\sourceset)$. Since $C_v(u_k) \cap V_{f}(C_v(u_k))=\emptyset$, it holds that $u_k \notin V_{f}(C_v(u_k))$, and hence by combining with claim (a),
we get that  $P_{s,u_k,v}[b_{s,u_k,v},u_k]\cap V_{f}(C_v(u_k))=\{b_{s,u_k,v}\}$.
Since by the proof of Lemma \ref{cl:clutone_suff}(b), $\phi_{s,t,v} \neq b_{s,u_k,v}$, hence
$Q_{s,t,v}[\phi_{s,t,v},u_k]\cap V_{f}(C_v(u_k))=\emptyset$.

Consider claim (d). By the above, $v$ occurs on the far segment $\pi(s, C_k) \setminus \{\LCA(s,C_k)\}$, hence $v \notin C_k$. Since $(\pi(s, C_k) \setminus C_k) \subseteq V_{f}(C_k)$, (d) holds.
\QED
The next observation is useful in our analysis.
\begin{observation}
\label{obs:stvno}
If $\phi_{s,t,v}$ satisfies (N1) and (N2), then there exists a vertex $x \in C_{v}(\phi_{s,t,v})$ satisfying that
$v \notin \pi(s,x)$.
\end{observation}
\Proof
Let $P_{\tau}=P_{s, t, v}$ and $\phi_\tau=\phi_{s,t,v}$.
By the uniqueness of the weight assignment $W$,
$P_{\tau}[s, \phi_{\tau}]=P_{s, \phi_{\tau},v}=SP(s, \phi_\tau, G \setminus \{v\},W)$.
%
Since $\phi_{\tau}$ satisfies (N2), it holds that the failing vertex $v$ occurs on $\pi^{near}(s, \phi_\tau)$, strictly \emph{below} (i.e., closer to $\phi_\tau)$ the least common ancestor $\LCA(s, C_v(\phi_{\tau}))$ on $\pi(s,\phi_\tau)$. Hence, there must exist a vertex $x \in C_v(\phi_{\tau})$ such that $v \notin \pi(s,x)$ (otherwise, if $v$ is shared by $\pi(s,u)$ for all cluster members $u$, then we end with contradiction to the definition of the least common ancestor $\LCA(s, C_v(\phi_{\tau}))$).
\QED
We proceed by showing correctness.
\begin{theorem}
\label{thm:correctnes_sw}
$H_8(S)$ is a $(8,S)$ FT-spanner.
\end{theorem}
\Proof
Let $H=H_8(S)$. It is required to show that $\dist(s, t, H \setminus \{v\}) \leq \dist(s, t, G \setminus \{v\})+8$ for every $(s,t) \in S \times V$ and $v \in V$.
By the analysis of Alg. \Conss\ (Lemma \ref{lem:correct4sw}),
it remains to consider the case of independent new-ending paths where $P_{s,t,v} \in \mathcal{P}^{far}_{indep}(t)$ for $t \in V_{\Delta}$.

Let $\tau$ be the iteration at which $P_{\tau}=P_{s,t,v}$ was considered to be added to the spanner at step (4.2), and let $\kappa=\Cost(P_{\tau})$ denote its cost.
Let $\phi_{\tau}$ be as defined in Def. \ref{def:phi} and recall that $Q_{\tau}=P_{\tau}[\phi_{\tau},t]$ is the candidate suffix to be bought by the procedure.
(In particular, $\Cost(P_{\tau})$ counts the number of edges on $Q_\tau \setminus E(G_{\tau})$.)
\paragraph{Case (1): $Q_{\tau}$ was bought.}
If $\phi_{\tau}$ did not satisfy neither properties (N1), (N2) (or both), then $P_{\tau}[s, \phi_\tau] \subseteq G_{\tau}$. Since
$P_{\tau}=P_{\tau}[s, \phi_{\tau}]\circ Q_{\tau}$ and $Q_{\tau}$ was added to the spanner, we get that $P_{\tau} \subseteq H \setminus \{v\}$.

\par It remains to consider the complementary case where  $\phi_{\tau}$ satisfies both (N1) and (N2).
By Obs. \ref{obs:stvno}, we get that
there exist $x \in C_{v}(\phi_{\tau})$ satisfying that
$v \notin \pi(s,x)$.

Consider the path $P=\pi(s,x) \circ (x, z_v(\phi_{\tau}),\phi_{\tau})$. By definition, $P \subseteq H \setminus \{v\}$ and by the existence of the intracluster path connecting $x$ and $\phi_\tau$ in $G \setminus \{v\}$, it holds that
$|P|=\dist(s,x, G \setminus \{v\})+2 \leq \dist(s,\phi_\tau, G \setminus \{v\})+4$.
Hence, letting $P'=P \circ Q_{\tau}$ (where $Q_\tau=P_{\tau}[\phi_\tau,t]$), since $Q_\tau \subseteq H \setminus \{v\}$, it holds that $P' \subseteq H \setminus \{v\}$ and $|P'|\leq |P_{\tau}|+4$, as required.

\paragraph{Case (2): $Q_{\tau}$ was not bought.}
Let $x \in C_v(\phi_{\tau})$ be defined as follows.
If $\phi_{\tau}$ satisfies both properties (N1) and (N2) of Def. \ref{def:phi}, then using Obs. \ref{obs:stvno}, let $x \in C_{v}(\phi_{\tau})$ be the vertex satisfying that $v \notin \pi(s,x)$.
Otherwise, if $\phi_{\tau}$ did not satisfy (N1) or (N2) (or both), let $x=\phi_{\tau}$. Note that in any case, it holds that $x, \phi_\tau \in C_v(\phi_\tau)$. We have the following.
\begin{lemma}
\label{cl:xpath}
$P_{s,x,v} \subseteq H \setminus \{v\}$.
\end{lemma}
\Proof
If $x=\phi_{\tau}$, then it implies that $\phi_\tau$ did not satisfy both of the properties (N1,N2). By Def. \ref{def:phi}, in such a case $\phi_\tau$ is the vertex incident to the first missing edge on $P_{s,t,v} \setminus E(G_\tau)$ and hence $P_{s,t,v}[s,x]=P_{s,x,v} \subseteq G_{\tau} \setminus \{v\}$.

Otherwise, if $x \neq \phi_{\tau}$, then $x \in C_v(\phi_\tau)$ and by the selection of $x$, $v \notin \pi(s,x)$. Hence, $P_{s,x,v}=\pi(s,x) \subseteq H \setminus \{v\}$.
\QED
Recall that $C_1(\tau)=C_v(\phi_\tau)$ and $C_2(\tau)=C_v(t)$. In addition, since $v \in \pi^{far}(s,t)$, it holds that $v \in V_f(C_2(\tau))$.

\begin{figure}[htbp]
\begin{center}
\includegraphics[scale=0.4]{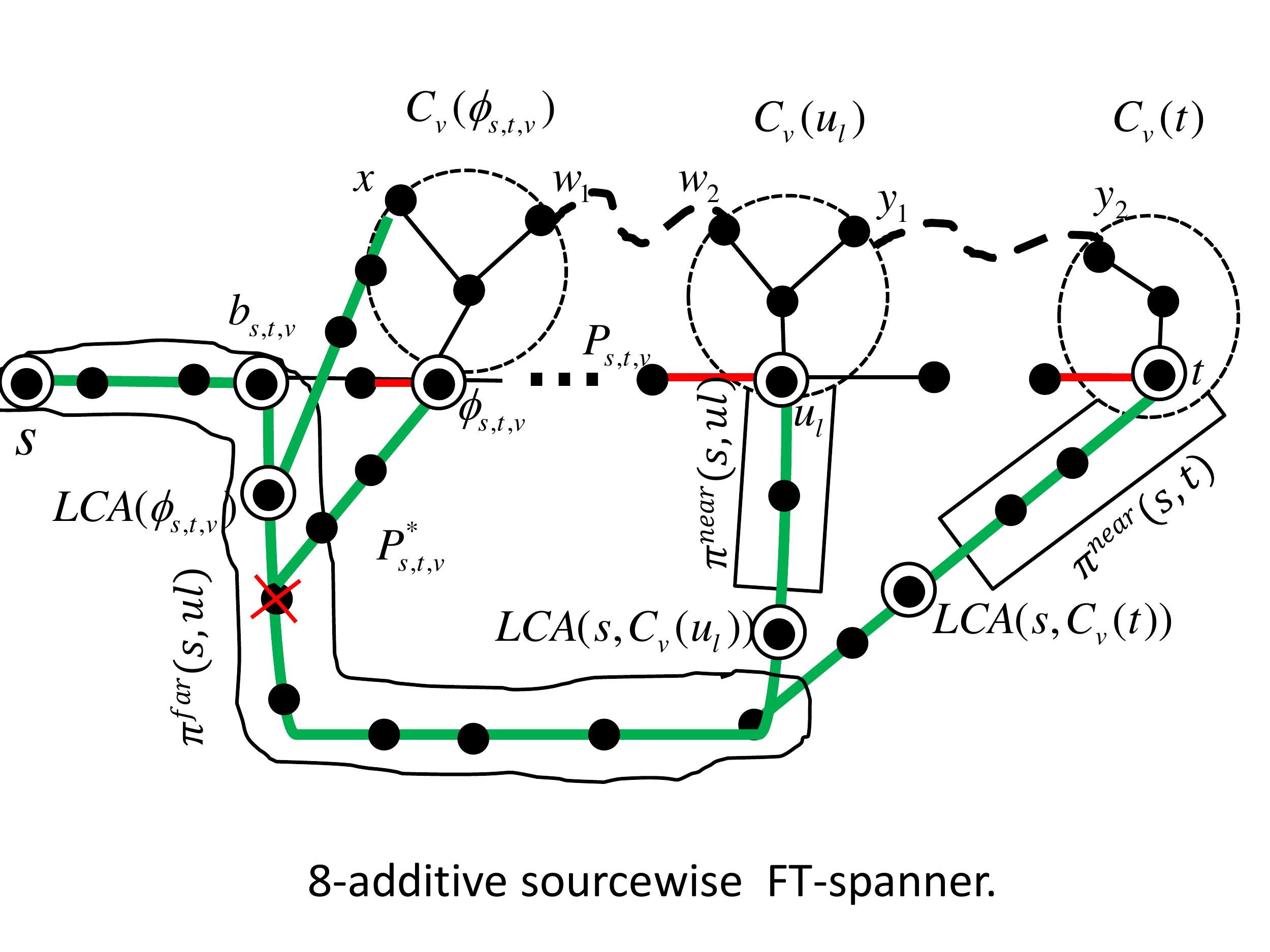}
\caption{Schematic illustration of the path-buying procedure of Alg.  \Consssix. The horizontal path is $P_{\tau}=P_{s,t,v}$ whose segment $Q_{s,t,v}=P_{s,t,v}[\phi_{s,t,v},t]$ was considered to be bought at time $\tau$. The green paths correspond to the shortest paths in $T_0(s)$. Red edges correspond to missing edges on $P_{s,t,v} \setminus E(G_{\tau})$. The vertex $\phi_{s,t,v}$ satisfies properties (N1) and (N2), hence it is incident to a missing edge and the failing vertex $v$ occurs on $\pi(s, \phi_{s,t,v})$ strictly below the LCA vertex $\LCA(\phi_{s,t,v})=\LCA(s, C_v(\phi_{s,t,v}))$. The vertex $x \in C_v(\phi_{s,t,v})$ satisfies that $v \notin \pi(s,x)$. The $s-t$ replacement path in $H \setminus \{v\}$ is given by traveling from $s$ to $x$ on $\pi(s,x)$ and then use the closest vertex pairs $w_1,w_2$ and $y_1,y_2$.
\label{fig:6add}}
\end{center}
\end{figure}

\paragraph{Case (2.1): $P_{\tau}$ is cheap.}
Since $Q_{\tau}$ was not added, Eq. (\ref{eq:short_pb}) did not hold and hence
\begin{equation}
\label{eq:ddcheap}
\dist(\phi_{\tau},t,P_{\tau})\geq \dist(C_1(\tau),C_2(\tau),G_{\tau} \setminus V_f(C_2(\tau)))~.
\end{equation}
Let $w_1 \in C_1(\tau)$ and
$w_2 \in C_2(\tau)$ be a closest pair satisfying that $\dist(w_1, w_2, G_{\tau} \setminus V_{f}(C_2(\tau)))=\dist(C_1(\tau), C_2(\tau), G_{\tau} \setminus V_{f}(C_2(\tau)))$.
Since the failing vertex $v$ is in $V_f(C_2(\tau))$, both auxiliary vertices $w_1$ and $w_2$ are in $G \setminus \{v\}$.
Consider the following $s-t$ path:
$P=P_0 \circ P_1 \circ P_2 \circ P_3$ where $P_0=P_{s,x,v}$, $P_1=[x, z_v(\phi_\tau), w_1]$, $P_2 \in SP(w_1, w_2, G_{\tau} \setminus V_{f}(C_2(\tau)))$, and $P_3=[w_2, z_v(t), t]$.
For an illustration see Fig.  \ref{fig:6add}. By Lemma \ref{cl:xpath}, $P_0 \subseteq H \setminus \{v\}$.
Note that since $x, w_1 \in C_v(\phi_\tau)$,
the path $P_1$ exists in $H \setminus \{v\}$.
Combining with the definitions of the vertices $z_v(x), z_v(t),w_1,w_2$, it holds that $P \subseteq H \setminus \{v\}$. So, it remains to bound the length of the path.
\begin{eqnarray}
\dist(s, t, H \setminus \{v\}) &\leq& |P_0|+|P_1|+|P_2|+|P_3| \nonumber
\\&=&
\dist(s,x,G \setminus \{v\})+\dist(w_1, w_2, G_{\tau} \setminus V_{f}(C_2(\tau)))+4 \nonumber
\\&\leq&
\dist(s,\phi_\tau,G \setminus \{v\}) \nonumber
\\&+& \dist(w_1, w_2, G_{\tau} \setminus V_{f}(C_2(\tau)))+6 \label{eq:pbsw_3}
\\&\leq&
\dist(s,\phi_\tau,G \setminus \{v\})+\dist(\phi_\tau, t, P_{\tau})+6~=|P_{\tau}|+6, \label{eq:pbsw_2}
\end{eqnarray}
where Eq. (\ref{eq:pbsw_3}) follows by the fact that $x,\phi_{\tau} \in C_v(\phi_{\tau})$ and since $G_{\Delta} \subseteq H$, it holds that the intracluster path $R=[x, z_v(\phi_\tau), \phi_\tau]$ exists in $G \setminus \{v\}$, Eq. (\ref{eq:pbsw_2}) follows by Eq. (\ref{eq:ddcheap}).

\paragraph{Case (2.2): $P_{\tau}$ is costly.}
Let
$U_{s,t,v}=\{u_{1}, \ldots, u_{\kappa'}\} \subseteq Q_{\tau}$ for $\kappa'=\lfloor \kappa/3 \rfloor \geq 1$ be as defined by Def. \ref{def:u}.
Since by Obs. \ref{obs:clust_prop}, the diameter of each cluster is 2, each $u_{k}\in U_{s,t,v}$ belongs to a distinct cluster $C_{k}=C_v(u_{k}) \in \mathcal{C}$.
Hence there are at least $\kappa'$ distinct clusters on $Q_{\tau}$.

\par A cluster $C_{k}=C_v(u_{k})$ is a \emph{contributor}
if adding $Q_\tau$ to $G_\tau$ improves either the $C_1(\tau)-C_{k}$ distance (i.e., $(C_1(\tau),C_k) \in \ValSet_1(P_{\tau})$)
or the $C_2(\tau)-C_{k}$ distance (i.e., $(C_k,C_2(\tau)) \in \ValSet_2(P_{\tau})$) in the corresponding appropriate graph.
%
Otherwise, $C_{k}$ is \emph{neutral}.
There are two cases to consider.
If all clusters are contributors (i.e., there is no neutral cluster) then all the $\kappa'$ clusters contribute to $\Value(P_\tau)$ (either with $C_1(\tau)$ or with $C_2(\tau)$ or both).
It then holds that $\Value(P_{\tau})\geq \kappa' \geq \Cost(P_\tau)/4$.  Hence, by Eq. (\ref{eq:cost_value}), we get a contradiction to the fact that the suffix $Q_\tau$ was not added to $G_\tau$.

In the other case, there exists at least one neutral cluster $C_{\ell}$ such that
\begin{eqnarray}
\label{eq:clust_correct}
\dist(C_1(\tau), C_{k} , \widehat{H}_{1}) &\leq&
\dist(\phi_\tau, u_{k} ,P_{\tau}) \mbox{~~and~~}
\\
\dist(C_{k} , C_2(\tau), \widehat{H}_{2}) &\leq& \nonumber
\dist(u_{k},t ,P_{\tau})~,
\end{eqnarray}
where $\widehat{H}_{1}=G_{\tau} \setminus V_{f}(C_{k})$ and $\widehat{H}_{2}=G_{\tau} \setminus V_{f}(C_2(\tau))$.
Let $w_1 \in C_1(\tau)$ and $w_2 \in C_{k}$ be the pair of vertices satisfying
$\dist(w_1,w_2 , \widehat{H}_{1})=\dist(C_1(\tau), C_{k} , \widehat{H}_{1})$.
In addition, let $y_1 \in C_k$ and $y_2 \in C_2(\tau)$ be the pair satisfying
$\dist(y_1,y_2 , \widehat{H}_{2})=\dist(C_{k} , C_2(\tau), \widehat{H}_{2})$.

Let $Q_1=[x, z_v(\phi_\tau), w_1]$, $Q_2=[w_2, z_v(u_{k}), y_1]$ and $Q_3=[y_2, z_v(t), t]$ be the intracluster paths in $C_1(\tau),C_k$ and $C_2(\tau)$ respectively. Note that by definition $x,w_1 \in C_v(\phi_\tau)$.

Since by Lemma \ref{lem:rep_new_hc}(c), $v \in V_{f}(C_k) \cap V_{f}(C_2(\tau))$, it also holds that $Q_1, Q_2, Q_3 \subseteq H \setminus \{v\}$.
Let $P'=P_0 \circ Q_1 \circ P_1 \circ Q_2 \circ P_2 \circ Q_3$ where $P_0=P_{s,x,v}$, $P_1 \in SP(w_1, w_2, \widehat{H}_{1})$ and $P_2 \in SP(y_1, y_2, \widehat{H}_{2})$. By Lemma \ref{cl:xpath}, $P_0 \subseteq H \setminus \{v\}$ and by the above explanation,
$P' \subseteq H \setminus \{v\}$. So, it remains to bound the length of the $s-t$ path $P'$.
\begin{eqnarray*}
\dist(s, t, H \setminus \{v\}) &\leq& |P'|=|P_0|+|P_1|+|P_2|+6
\\&=&
\dist(s,x, G\setminus \{v\})+\dist(w_1, w_2, \widehat{H}_{1})+\dist(y_1,y_2, \widehat{H}_{2})+6
\\& \leq &
\dist(s,\phi_\tau, G\setminus \{v\})+\dist(w_1, w_2, \widehat{H}_{1})+\dist(y_1,y_2, \widehat{H}_{2})+8
\\& = &
\dist(s,\phi_\tau, G\setminus \{v\})+\dist(C_1(\tau), C_{k}, \widehat{H}_{1})
\\&+&
\dist(C_{k}, C_2(\tau), \widehat{H}_{2})+8
\\& \leq &
\dist(s,\phi_\tau, G\setminus \{v\})+\dist(\phi_\tau, u_{k}, P_{\tau})+\dist(u_{k}, t, P_{\tau})+8
\\&=&
|P_{s,t,v}|+8~,
\end{eqnarray*}
where the first inequality follows by the fact that $x,\phi_\tau \in C_v(\phi_\tau)$ and hence the intraclusrer path $R=[x, z_v[\phi_\tau], \phi_\tau]$ exists in $G \setminus \{v\}$ and last inequality follows Eq. (\ref{eq:clust_correct}). The claim follows.
\QED

Finally, we turn to bound the size of $H_8(S)$ and claim the following.
\begin{lemma}
\label{lem:sourcew_size}
$|E(H_8(S))|=\widetilde{O}(\max\{|S| \cdot n, n^{2}/|S|^2\})$.
\end{lemma}
\Proof
Let $H=H_8(S)$.
By the size-analysis of Alg. \Conss, it remains to bound the number of edges added due to the path-buying procedure of step (4.2).

Let $\mathcal{B} \subseteq \mathcal{P}^{far}_{indep}$ be the set of paths corresponding to the paths segments that were bought in the path-buying phase.
For every ordered  pair of clusters, $C_1, C_2 \in \mathcal{C}$ let $\mathcal{B}(C_1, C_2)=\{ P_{\tau} \in \mathcal{B} ~\mid~ (C_1, C_2) \in \ValSet(P_{\tau})\}$.

Clearly, $\mathcal{B}=\bigcup_{C_1, C_2 \in \mathcal{C}} \mathcal{B}(C_1, C_2)$. We next claim that since the diameter of each cluster is small, it holds that the cardinality of each subset $\mathcal{B}(C_1, C_2)$ is small as well.
\begin{lemma}
$|\mathcal{B}(C_1, C_2)| \leq 5$ for every $C_1, C_2 \in \mathcal{C}$.
\end{lemma}
\Proof
Fix $C_1, C_2 \in \mathcal{C}$ and let $\mathcal{B}(C_1, C_2)=\{P_{\tau_1}, \ldots, P_{\tau_N}\}$ be sorted according to the time $\tau_k$ their segment $Q_{\tau}$ was added to the spanner, for every $k \in \{1, \ldots, N\}$ where $N=|\mathcal{B}(C_1, C_2)|$. Let $P_{\tau_k}=P_{s_k,t_k,v_k}$. Let $p_k, q_k \in P_{\tau_k}$ be such that $p_k$ is closer to the source $s_k$, and
$C_{v_k}(p_k)=C_1$ and $C_{v_k}(q_k)=C_2$.

Recall that $\phi_{\tau_k}$ is the first vertex of $Q_{\tau_k}$ (see Def. \ref{def:phi}).
Let $C_\ell=C_{v_k}(u_\ell)$ be the cluster of $u_{\ell}$ for every $u_\ell \in U_{s_k,t_k,v_k}$ (see Def. \ref{def:u}).
By Obs. \ref{obs:clust_prop}, it holds that $C_{\ell} \neq C_{\ell'}$ for every $u_{\ell},u_{\ell'} \in U_{s_k,t_k,v_k}$.
Recall that for every $u_\ell \in  U_{s_k,t_k,v_k}$, $P_{s,u_{\ell},v}=P_{s,t,v}[s,u_{\ell}]$.
Since $\LastE(P_{s,u_\ell,v}) \notin E^{local}$, it holds that $v \notin \{z_1(u_{\ell}), \LCA(s, C_1(u_{\ell}))\}$. Combining that with the fact that $u_\ell \in U_{s_k,t_k,v_k}$ did not satisfy property (N2) (see Def. \ref{def:phi} and Def. \ref{def:u}), we conclude that $v_k \in \pi^{far}(s_k, u_\ell)$.
Since $q_k \in U_{s_k,t_k,v_k} \cup \{t_k\}$, using Lemma \ref{lem:rep_new_hc}(c), it holds that
\begin{equation}
\label{eq:pb_failedvsw}
v_k \in V_{f}(C_2) \mbox{~for every~} k \in \{1,\ldots,N\}~,
\end{equation}
and by Lemma \ref{lem:rep_new_hc}(b),
\begin{equation}
\label{eq:pb_notinhcsw}
P_{\tau_k}[p_k, q_k] \subseteq Q_{\tau_k}[p_k, q_k] \subseteq G \setminus V_{f}(C_2)~.
\end{equation}

Since $C_1=C_{v_k}(p_k)$ and $C_2=C_{v_k}(q_k)$, for every $k \in \{1, \ldots, N\}$, it holds that $z_{v_{1}}(p_1)=...=z_{v_{N}}(p_N)$ and also
$z_{v_{1}}(q_1)=...=z_{v_{N}}(q_N)$. Hence, letting $z_1=z_{v_{1}}(p_1)$ and $z_2=z_{v_{1}}(q_1)$, it holds that
\begin{equation}
\label{eq:clustercenternotfullsw}
z_1,z_2 \notin \{v_1, \ldots, v_N\}.
\end{equation}

Denote
$$X_k=\dist(p_k, q_k, G_{\tau_{k+1}} \setminus V_{f}(C_2)).$$
We now show that $X_k<X_{k-1}$ for every $k \in \{2, \ldots N\}$.

Each time a path segment $Q_{\tau_k}$ is purchased at time $\tau_k$, it implies that
\begin{eqnarray}
X_k &\leq&
\dist(p_k, q_k, P_{\tau_k} \setminus V_{f}(C_2)) \label{eq:adub_111}
\\&=&
\dist(p_k, q_k,P_{\tau_k}) \label{eq:adub_121}
\\&<&
\dist(C_{1}, C_{2}, G_{\tau_{k}} \setminus V_{f}(C_2)) \label{eq:adub_131}
\\&\leq&
X_{k-1}~,
\label{eq:adub_4}
\end{eqnarray}
where Eq. (\ref{eq:adub_111}) follows by the fact that $P_{\tau_k}[p_k, q_k] \subseteq Q_{\tau_k} \subseteq G_{\tau_{k+1}}$, Eq. (\ref{eq:adub_121}) follows by Eq. (\ref{eq:pb_notinhcsw}),
Eq. (\ref{eq:adub_131}) follows by the fact that $Q_{\tau_k}$ was bought and by Eqs. (\ref{eq:contvs1}) and (\ref{eq:contvs2}), and Eq. (\ref{eq:adub_4}) follows by the fact that $p_{k-1} \in C_1$ and $q_{k-1} \in C_2$.

Therefore,
we have that
\begin{equation}
\label{eq:up_srtect11}
X_N \leq  X_1-(N-1)~.
\end{equation}
Conversely, we have that
\begin{eqnarray}
X_N
&\geq&
\dist(p_{N}, q_N, G \setminus \{v_1, \ldots, v_N\}) \label{eq:up_t011}
\\
&\geq&
\dist(p_1, q_1, G \setminus \{v_1, \ldots, v_N\})-4 \label{eq:up_t021}
\\
& =&
\dist(p_1, q_1, P_{\tau_1})-4 \label{eq:up_t031}
= \dist(p_1, q_1, P_{\tau_1} \setminus V_{f}(C_2))-4
\\& \geq &
X_1-4~, \label{eq:up_t041}
\end{eqnarray}
where Eq. (\ref{eq:up_t011}) follows as $G_{\tau_{N+1}} \subseteq G$ and by Eq. (\ref{eq:pb_failedvsw}), $\{v_1, \ldots, v_N\} \subseteq V_{f}(C_2)$.
To see Eq. (\ref{eq:up_t021}), note that $p_{1},p_N \in C_{1}$ and $q_1,q_N \in C_{2}$ and by Obs. \ref{obs:clust_prop}(3) the diameter of the cluster is 2.
It remains to show that the intracluster paths $R_1=[p_1, z_1, p_N], R_2=[q_1, z_2, q_N]$ exist in the surviving graph $G \setminus \{v_1, \ldots, v_N\}$. This holds since by Eq. (\ref{eq:clustercenternotfullsw}),
$z_1,z_2 \notin \{v_1, \ldots, v_N\}$, and by Eq. (\ref{eq:pb_failedvsw}) and (\ref{eq:pb_notinhcsw}).
Eq. (\ref{eq:up_t031}) follows by Eq. (\ref{eq:pb_notinhcsw}).
%
%
Finally, Eq. (\ref{eq:up_t041}) follows by the fact that $P_{\tau_1}[p_1,q_1]\subseteq Q_{\tau_1}$ was added at step $\tau_1$, hence $P_{\tau_1}[p_1,q_1] \subseteq G_{\tau_{2}}$. By combining with Eq. (\ref{eq:up_srtect11}), we get that $N \leq 5$. The claim follows.
\QED
Finally, since for every path $P \in \mathcal{B}$, it holds that $\Cost(P) \leq 4 \cdot \Value(P)$, we get that
\begin{eqnarray*}
|E(G_{\tau'}) \setminus E(G_0)|&=& \sum_{P \in \mathcal{B}} \Cost(P)
\leq
4 \sum_{P \in \mathcal{B}} \Value(P)
\leq
4 \sum_{C_1, C_2 \in \mathcal{C}} |\mathcal{B}(C_1, C_2)|
\\& \leq &
O(|\mathcal{C}|^2)=\widetilde{O}((n/|S|)^2)~.
\end{eqnarray*}
where the last equality follows by the fact that there are $|\mathcal{C}|=|Z|=\widetilde{O}(n/|S|)$ clusters. The claim follows.
\QED
\paragraph{Additive stretch $6$ (for all pairs).}
\label{sec:6add}
Thm. \ref{thm:6stretch} follows immediately by  Lemma \ref{lem:h4}. This should be compared with the single source additive FT-spanner $H_4(\{s\})$ of \cite{PPFTBFS14} and the (all-pairs, non FT) 6-additive spanner, both with $\widetilde{O}(n^{4/3})$ edges.

\end{document}